\documentclass[12pt,compress,nosort,hyper]{JHEP3}
\usepackage{cite,amsmath,exscale,amssymb,latexsym,axodraw}
\usepackage[dvips]{graphics}

\def\Year{\expandafter\eatPrefix\the\year}
\newcount\hours \newcount\minutes
\def\monthname{\ifcase\month\or
January\or February\or March\or April\or May\or June\or July\or
August\or September\or October\or November\or December\fi}
\def\shortmonthname{\ifcase\month\orx
Jan\or Feb\or Mar\or Apr\or May\or Jun\or Jul\or
Aug\or Sep\or Oct\or Nov\or Dec\fi}

\def\TimeStamp{\hours\the\time\divide\hours by60%
\minutes -\the\time\divide\minutes by60\multiply\minutes by60%
\advance\minutes by\the\time%
${\rm \shortmonthname}\cdogv paper20t   \if\day<10{}0\fi\the\day\cdot   \the\year
\qquad\the\hours:\if\minutes<10{}0\fi\the\minutes$}

\newskip\humongous \humongous=0pt plus 1000pt minus 100pt
\def\caja{\mathsurround=0pt}
\def\eqalign#1{\,\vcenter{\openup1\jot \caja
       \ialign{\strut \hfil$\displaystyle{##}$&$
        \displaystyle{{}##}$\hfil\crcr#1\crcr}}\,}
\newif\ifdtup

\newcounter{eqnumber}[section]
\renewcommand{\theeqnumber}{\thesection.\arabic{eqnumber}}
\def\equn{\refstepcounter{eqnumber}
\eqno({\rm \theeqnumber})
}

\def\eqn#1{eq.~(\ref{#1})}

\def\sec#1{section~{\ref{#1}}}

\newbox\charbox
\newbox\slabox
\def\s#1{{      
        \setbox\charbox=\hbox{$#1$}
        \setbox\slabox=\hbox{$/$}
        \dimen\charbox=\ht\slabox
        \advance\dimen\charbox by -\dp\slabox
        \advance\dimen\charbox by -\ht\charbox
        \advance\dimen\charbox by \dp\charbox
        \divide\dimen\charbox by 2
        \raise-\dimen\charbox\hbox to \wd\charbox{\hss/\hss}
        \llap{$#1$}
}}

\def\spa#1.#2{\left\langle#1\,#2\right\rangle}
\def\spb#1.#2{\left[#1\,#2\right]}
\def\lor#1.#2{\left(#1\,#2\right)}

\catcode`@=11  

\def\Slash#1{\hskip 0.05 cm \slash\hskip -0.22 cm #1}

\def\Tr{\, {\rm Tr}}

\def\eps{\epsilon}

\def\Ord{{\cal O}}

\def\e{\epsilon}

\def\textcolor#1{{}}

\def\la{\langle}
\def\ra{\rangle}
\def\oneloop{{\rm one \mbox{-} \rm loop}}

\def\lsl{\not{\hbox{\kern-2.3pt $\ell$}}}
\def\ksl{\not{\hbox{\kern-2.3pt $k$}}}

\def\spa#1.#2{\left\langle#1\,#2\right\rangle}
\def\spb#1.#2{\left[#1\,#2\right]}
\def\lor#1.#2{\left(#1\,#2\right)}

\def\sand#1.#2.#3{%
  \left\langle\smash{#1}{\vphantom1}\right|{#2}%
  \left|\smash{#3}{\vphantom1}\right\rangle}
\def\sandp#1.#2.#3{%
  \left\langle\smash{#1}{\vphantom1}^{-}\right|{#2}%
  \left|\smash{#3}{\vphantom1}^{+}\right\rangle}
\def\sandpp#1.#2.#3{%
  \left\langle\smash{#1}{\vphantom1}^{+}\right|{#2}%
  \left|\smash{#3}{\vphantom1}^{+}\right\rangle}
\def\sandmm#1.#2.#3{%
  \left\langle\smash{#1}{\vphantom1}^{-}\right|{#2}%
  \left|\smash{#3}{\vphantom1}^{-}\right\rangle}
\def\sandpm#1.#2.#3{%
  \left\langle\smash{#1}{\vphantom1}^{+}\right|{#2}%
  \left|\smash{#3}{\vphantom1}^{-}\right\rangle}
\def\sandmp#1.#2.#3{%
  \left\langle\smash{#1}{\vphantom1}^{-}\right|{#2}%
  \left|\smash{#3}{\vphantom1}^{+}\right\rangle}

\def\Atree{A^{\rm tree}}

\def\Lzz{\mathop{\hbox{\rm L}}\nolimits_2}
\def\Lz{\mathop{\hbox{\rm L}}\nolimits_0}
\def\Kz{\mathop{\hbox{\rm K}}\nolimits_0}

\def\L{\left(}\def\R{\right)}

\def\L{\left(}\def\R{\right)}

\def\Lzi{\mathop{\hbox{\rm L}}\nolimits_i}

\def\tree{{\rm tree}}

\def\Gr{{\rm Gr}}

\def\NeqFour{{\cal N} = 4}
\def\NeqOne{{\cal N} = 1}
\def\NeqZero{{\cal N} = 0}
\def\BRQ#1#2#3{\la #1|{#2}|#3\ra}
\def\BBRQ#1#2#3{[ #1|{#2}|#3\ra}

\def\Lzi{\mathop{\hbox{\rm L}}\nolimits_i}

\def\Fact{{\cal F}}
\def\inlimit^#1{\buildrel#1\over\llongrightarrow}
\def\llongrightarrow{%
\relbar\mskip-0.5mu\joinrel\mskip-0.5mu\relbar
     \mskip-0.5mu\joinrel\longrightarrow}

\title{Recursive Calculation of One-Loop QCD Integral Coefficients}

\author{
Zvi~Bern\\
Department of Physics\\
University of California at Los Angeles \\
Los Angeles, CA 90095, USA}
\author{
N.~E.~J.~Bjerrum-Bohr,
David~C.~Dunbar and
Harald~Ita\\
Department of Physics \\
University of Wales Swansea \\
Swansea, SA2 8PP, UK}
\preprint{
hep-ph/0507019\\
UCLA-TEP-/05/19\\
SWAT-05-434}

\abstract{We present a new procedure, using on-shell recursion, to
determine coefficients of integral functions appearing in the one-loop
scattering amplitudes of gauge theories, including QCD.  With this
procedure, coefficients of integrals, including bubbles and triangles,
can be determined without resorting to integration. We give criteria
for avoiding spurious singularities and boundary terms that would
invalidate the recursion.  As an example where the criteria are
satisfied, we obtain all cut-constructible contributions to the
one-loop $n$-gluon scattering amplitude,
$A_n^{\oneloop}(\ldots---+++\ldots)$\,, with split-helicity from an
$\NeqOne$ chiral multiplet and from a complex scalar.  Using the
supersymmetric decomposition, these are ingredients in the
construction of QCD amplitudes with the same helicities.  This method
requires prior knowledge of amplitudes with sufficiently large numbers
of legs as input. In many cases, these are already known in compact
forms from the unitarity method. }


\begin{document}


\section{Introduction}

With the approach of the Large Hadron Collider, perturbative QCD
computations will play an ever more prominent role in helping to
unravel physics beyond the Standard Model. Computations at tree-level
provide a first step for providing theoretical guidance. For reliable
quantitative computations, at a minimum, next-to-leading order
calculations are required. Over the past decade important strides have
been taken in producing new next-to-leading order calculations to various
standard model processes in pure QCD, heavy quark physics, and mixed
QCD/electroweak processes. (See, for example, refs.~\cite{QCDReviews}
for recent summaries.) However, much more needs to be done to take
full advantage of the forthcoming data.

In this paper we discuss one aspect of this problem: the
computation of coefficients of integral functions appearing in
one-loop scattering amplitudes with large numbers of external legs. In
a very recent paper~\cite{BDKSix}, an improved unitarity-factorisation
bootstrap approach~\cite{ZFourPartons} was presented for obtaining
complete amplitudes, including the rational function terms.  As
examples, one-loop six- and seven-point QCD amplitudes with two
colour-adjacent negative helicities and the rest positive were
computed. These rational function pieces are trivial in supersymmetric
theories, but presented a bottleneck to further progress in QCD loop
calculations.  With this obstruction mitigated, it is worthwhile to
re-examine the cut containing pieces, as we do here, to see if further
computational improvements may be found, especially in the
light of recent
progress.

The recent advances have been stimulated by Witten's proposal
of a `weak-weak' duality between a string theory and $\NeqFour$ gauge
theory~\cite{Witten2003nn} (generalising a previous description of the
simplest gauge theory amplitudes by Nair~\cite{Nair1988bq}).  (For a
recent review of these developments, see ref.~\cite{Cachazo2005ga}.)
Some of these efforts have been focused on analysing amplitudes from the
viewpoint of $\NeqFour$ super-Yang-Mills theory as a
topological string theory in twistor
space~\cite{topstring}. There has at the same time been rapid progress
for calculating amplitudes on the field theory side. This has led to a
variety of calculations and insights into the structure of tree
amplitudes~\cite{CSW,Trees,CSWmatter,TreeRecursion,BCFW,
TreeRecursionResults,BBSTgravity,SplitHelicity}
and supersymmetric one-loop
amplitudes~\cite{OneloopMHVVertices,CachazoHolomorphic,BCFHolomorphic,
BeDeDiKo,OneloopMHVVerticesB,BBBDDNEq1,BBDP,BrittoUnitarity,OneloopMHVVerticesC,
BDKn,BeBbDu,BDKrecursionA,BBDPSQCD,BrittoSQCD,BDKrecursionB,Bidder2005in,
Brandhuber2005QCD}. The novel organisation of string amplitudes has led to
reorganisations of previously computed loop scattering
amplitudes~\cite{StringBased,FirstQuantised}.
There have also been some promising steps for
massive theories~\cite{CSWmassive,Badger2005zh} and at
multiloops~\cite{BRY}.  The conclusion of these studies is that
gauge theory helicity amplitudes are extremely
simple, especially when compared to expectations.  This
simplicity has been the key to rapid progress.

For the pieces of one-loop amplitudes containing cuts, the unitarity
method~\cite{BDDKa,BDDKb,ZFourPartons,TwoLoopSplitting,
BeDeDiKo,BrittoUnitarity,BDKn}, due to Dixon, Kosower and two of the
authors, provides an effective means of computation.  This method has
been applied successfully in a variety of calculations in
QCD~\cite{BernMorgan,ZFourPartons,QCDUnitarity,TwoLoopSplitting} and
in supersymmetric theories~\cite{BDDKa,BDDKb,BRY,BeDeDiKo,BDKn}.  A
recent improvement~\cite{BrittoUnitarity} to the unitarity method
making use of generalised
unitarity~\cite{ZFourPartons,TwoLoopSplitting}, allows coefficients of
box integrals to be determined without explicit integration.  Other
related methods include the very beautiful application of MHV
vertices~\cite{CSW} to loop
calculations~\cite{OneloopMHVVertices,OneloopMHVVerticesB} and the
use~\cite{CachazoHolomorphic,BCFHolomorphic,BBBDDNEq1} of the
holomorphic anomaly~\cite{HolomorphicAnomaly} to
freeze~\cite{HolomorphicFreeze,CachazoHolomorphic} the cut integrals.
The unitarity method can also be used to determine complete
non-supersymmetric amplitudes, including the rational function terms,
by applying $D$-dimensional unitarity; a number of examples for four
external particles, and up to six for special helicity configurations,
have been worked out this way~\cite{BernMorgan,DimShift,QCDUnitarity,
TwoLoopSplitting,Brandhuber2005QCD}.

We make use of the decomposition of one-loop QCD $n$-gluon
scattering amplitudes into contributions originating from $\NeqFour$
and $\NeqOne$ supersymmetric multiplets and those originating from a
scalar loop~\cite{FiveGluon,BDDKa,BDDKb}.  Although supersymmetric multiplets
contain a more complicated particle content, it is simpler to
calculate the QCD amplitudes in this fashion; each of the pieces
is naturally distinguished by their differing analytic structures.
The simplest
contribution is the most supersymmetric multiplet, {\it i.e.}, the
contribution of an $\NeqFour$ super Yang-Mills multiplet.  The
$\NeqFour$ amplitudes can be expressed as scalar box integral
functions with rational coefficients~\cite{BDDKa} and these
coefficients have been evaluated in a closed form for the case with
maximally helicity violating (MHV) helicity configurations where two
gluons are of negative helicity and the rest of positive
helicity~\cite{BDDKa,OneloopMHVVertices} and for the case of next-to-MHV (NMHV)
amplitudes~\cite{BDDKb,BCFHolomorphic,BeDeDiKo,BDKn}.  (At tree-level
the corresponding helicity amplitudes are the Parke-Taylor
amplitudes~\cite{ParkeTaylor}.)  Using supersymmetric Ward identities
for one-loop $\NeqFour$ NMHV box coefficients, expressions involving
external gluino and quark legs have also been presented in
ref.~\cite{Bidder2005in}. Results in the case of an $\NeqOne$ chiral
multiplet have also been computed including, all-$n$ one-loop MHV
and all one-loop gluon six-point amplitudes~\cite{BDDKb,OneloopMHVVerticesB,
BBBDDNEq1,BBDP,BBDPSQCD,BrittoSQCD}.
In the scalar loop case the cut parts of the MHV amplitudes are
known~\cite{BDDKa,OneloopMHVVerticesC} for an arbitrary number of
legs.  Very recently an explicit form of the complete amplitude,
including rational functions has been presented at six-points for the
case of nearest neighbouring negative helicities in the colour
ordering~\cite{BDKSix}.

In this paper we focus on the question of improving calculational
methods for determining the coefficients of the integrals functions
containing the cut constructible parts of amplitudes. As explained in
ref.~\cite{BrittoUnitarity}, by considering quadruple cuts in the
unitarity method one can determine the coefficients of box integrals
rather directly, without the needing to perform any integral
reductions or integrations.  However, in evaluations of
$\NeqOne$ super-QCD or in non-supersymmetric QCD one encounters bubble
or triangle tensor integrals, which have to be directly
integrated; in general, experience
shows that it is best to avoid direct integration, whenever
possible. Even with this complication, as mentioned above, a wide
variety of amplitudes in the $\NeqOne$ theory have already been
computed. However, one may wonder if further improvements can be
found.

The approach we take here is based on recursion.  Recursion relations
have been very successfully used in QCD, starting with the
Berends-Giele recursion relations~\cite{BerendsGiele}.  The
Berends-Giele recursion relations have also been applied in some
special cases at loop level~\cite{Mahlon}. Recently at tree-level a
new recursive approach using on-shell amplitudes, but with complex
momenta has been written down by Britto, Cachazo and
Feng~\cite{TreeRecursion}.  This has led to compact expressions for
tree-level gravity and gauge theory
amplitudes~\cite{TreeRecursionResults,BBSTgravity,SplitHelicity}.
This approach appears to work very generally, including massive
amplitudes~\cite{Badger2005zh}.

An elegant and simple proof of the on-shell recursion relations has
been given by Britto, Cachazo, Feng and Witten~\cite{BCFW}.  Their
proof is actually quite general, and applies to any rational function
of the external spinors satisfying certain scaling and factorisation
properties.  Although the recursion relations emerged from loop
calculations~\cite{BeDeDiKo,RSVTreeRecursion,TreeRecursion}, the proof
does not extend straightforwardly to loop level because of a number of
issues.  The most obvious difficulty is that loop amplitudes
contain branch cuts which violate the assumption that the amplitude
consists of a sum of simple poles, used to prove the recursion.
Moreover, there are no theorems as yet on the factorisation properties
of loop amplitudes with complex momenta.  Indeed there are a set of
`unreal pole'~\cite{BDKrecursionA,BDKrecursionB} contributions that
must be taken into account, but whose nature is not yet fully understood.

Initial steps in extending on-shell recursion to computations of loop
amplitudes were taken in refs.~\cite{BDKrecursionA,BDKrecursionB},
where compact expressions were found for all remaining loop amplitudes
in QCD which are finite and composed purely of rational functions.
This however, still left the question of how one might compute cut
containing one-loop amplitudes in QCD via on-shell recursion.  As
mentioned above, a new approach has very recently been devised for
doing so~\cite{BDKSix}.  The method systematises an earlier
unitarity-factorisation approach first applied to computing the
one-loop amplitudes required for $Z \rightarrow 4$ jets and $p p
\rightarrow W + 2$ jets at next-to-leading order in the QCD
coupling~\cite{ZFourPartons}.  With this method one assumes that the
cut containing parts have already been computed using the unitarity or
related methods. The new approach then provides a systematic means via
on-shell recursion for obtaining the rational function pieces,
accounting for overlaps with the cut containing terms.

In this paper, we address the complementary question of whether one can
also construct on-shell recursion relations for the cut containing
pieces.  We sidestep the issue of applying on-shell recursion in the
presence of branch cuts by considering recursion for the rational
coefficients of integral functions and not for full
amplitudes.  The key ingredients that will enter our construction are:
\begin{itemize}

\item The universal factorisation properties that one-loop amplitudes
must satisfy, allowing us to deduce the factorisation properties of
integral coefficients~\cite{BDDKa,BDDKb,BernChalmers}.

\item The structure of spurious singularities that can appear in
dimensionally regularised one-loop amplitudes~\cite{Integrals,BernChalmers}.

\item The basis of integrals in which dimensionally regularised
amplitudes may be expressed~\cite{EarlierIntegrals,Integrals}.

\item The proof of the tree-level on-shell recursion relations due to
      Britto, Cachazo, Feng and Witten~\cite{BCFW}.

\item Prior calculations of integral coefficients that may be used as
starting points in the recursive procedure.  In this sense our procedure
dovetails very nicely with unitarity methods, which provide the
necessary inputs to the recursion.

\end{itemize}

We will present simple sufficiency criteria to guarantee that
a recursion on the coefficients is valid, with no boundary terms
or spurious singularities affecting the recursion.

As a non-trivial example of our recursion procedure we
will consider one-loop $n$-gluon ``split-helicity''
amplitudes of the form,
$$
A(1^-,2^-,\cdots l^-,(l+1)^+,\cdots, n^+)\,,
\equn\label{SplitHelicityAmplitude}
$$ where all negative helicity gluons are colour-adjacent.  At tree
level, amplitudes with this helicity configuration were computed in
ref.~\cite{SplitHelicity}.  In this paper we compute two
components of the supersymmetric decomposition of one-loop QCD
amplitudes for this configuration: The $\NeqOne$ chiral multiplet
contribution and the non-rational or ``cut-constructible'' part of the
scalar amplitude.  For three negative helicities $n$-point
amplitudes of the split-helicity form,
the $\NeqFour$~\cite{BDKn} and $\NeqOne$~\cite{BBDPSQCD} amplitudes
were derived earlier.  We will present the cut parts of the scalar loop
contributions, leaving the rational function parts as the final piece
to be computed.


\section{Notation}

Throughout this paper we employ colour-ordered
amplitudes. At tree level, the full amplitudes decomposes
in the following way
$$\hspace{0.1cm}\eqalign{
{\cal A}_n^{\rm tree}(1,2,\ldots, n) &\; = \; g^{(n-2)}
\hspace{-0.1cm}\sum_{\sigma \in S_n/Z_n}
\hspace{-0.2cm}{\rm Tr}\left( T^{a_{\sigma(1)}}
T^{a_{\sigma(2)} }\cdots  T^{a_{\sigma(n)}} \right)
 A_n^{\rm tree}(\sigma(1), \sigma(2),\ldots, \sigma(n)) \,,
\cr}\equn\label{TreeAmplitudeDecomposition}
$$
where $S_n/Z_n$ is the set of all permutations, but with cyclic
rotations removed. The $T^{a_i}$ are fundamental representation
matrices for the Yang-Mills gauge group $SU(N_c)$\,, normalised so that
$\Tr(T^aT^b) = \delta^{ab}$\,.
(For more detail on the tree and one-loop colour ordering of
gauge theory amplitudes see refs.~\cite{TreeReview,Colour}.)

For one-loop amplitudes of adjoint representation particles in the
loop, one may perform a colour decomposition similar to the tree-level
decomposition~(\ref{TreeAmplitudeDecomposition})~\cite{Colour}.  In
this case there are two traces over colour matrices and the result
takes the form,
$$
{\cal A}_n^{\oneloop}\L \{k_i,a_i\}\R \; = \;
g^n \,\sum_{c=1}^{\lfloor{n/2}\rfloor+1}
      \sum_{\sigma \in S_n/S_{n;c}}
     \Gr_{n;c}\L \sigma \R\,A_{n;c}^{}(\sigma)\,,
\label{ColourDecomposition}\equn
$$
where ${\lfloor{x}\rfloor}$ is the largest integer less than or equal to $x$\,.
The leading colour-structure factor,
$$
\Gr_{n;1}(1) \; = \; N_c\ \Tr\L T^{a_1}\cdots T^{a_n}\R \,,\equn
$$
is just $N_c$ times the tree colour factor, and the subleading colour
structures ($c>1)$ are given by,
$$
\Gr_{n;c}(1) \; = \; \Tr\L T^{a_1}\cdots T^{a_{c-1}}\R\,
\Tr\L T^{a_c}\cdots T^{a_n}\R \,.\equn
$$
$S_n$ is the set of all permutations of $n$ objects
and $S_{n;c}$ is the subset leaving $\Gr_{n;c}$ invariant.
Once again it is convenient to use $U(N_c)$ matrices; the extra $U(1)$
decouples~\cite{Colour}.
The contributions with fundamental representation quarks can be
obtained from the same partial amplitudes, except that sum runs only
over the $A_{n;1}$ and the overall factor of $N_c$ in $\Gr_{n;1}$ is
dropped.

For one-loop amplitudes the subleading in colour amplitudes
$A_{n;c}$\,,   $c > 1 $\,,
may be obtained from summations of permutations
of the leading in colour amplitude~\cite{BDDKa},
hence, we need only focus on the leading in colour amplitude $A_{n;1}$\,,
which we will generally abbreviate to $A_n$\,,
and use this
relationship to generate the full amplitude if required.

We will represent amplitudes employing the spinor-helicity
formalism~\cite{SpinorHelicity}.  Here amplitudes are expressed in
terms of spinor inner-products
$$
\spa{j}.{l} \;\equiv\; \langle j^- | l^+ \rangle \; = \;
\overline{u}_-(p_j) u_+(p_l)\,, \hskip 2 cm \spb{j}.{l} \;\equiv\;
\langle j^+ | l^- \rangle \; = \; \overline{u}_+(p_j) u_-(p_l)\,,
\equn \label{spinorproddef}
$$%
where $u_\pm(p)$ is a massless Weyl spinor with momentum $p$ and positive
or negative chirality.  With the normalisations used here,
$\spb{i}.{j} = {\rm sign}(p_i^0 p_j^0)\spa{j}.{i}^*$
so that,
$$
\spa{i}.{j} \spb{j}.{i} \;\equiv\; 2 p_i \cdot p_j \; = \; s_{ij}\,, \equn
$$
where
$$s_{i,i+1} \;\equiv\; K_{i,i+1}^2 \; = \; (p_i \;+\; p_{i+1})^2\ \ \ \ {\rm and}\ \ \
\ t_{i,j} \;\equiv\; K_{i,j}^2 \; = \; (p_i\; +\; \ldots \;+\; p_j)^2\,,\equn
$$ for generic
cyclic ordered sets of external momenta $K_{i,j}=p_i +\ldots +
p_j$ counting from leg $i$ to end leg $j$\,.
Note that $\spb{i}.{j}$ defined in this way differs by an
 overall sign from the
notation commonly used in twistor-space studies~\cite{Witten2003nn}. As in the twistor-space studies we define
$$
\lambda_i \;=\;  u_+(p_i)
\; , \;\;\;
\tilde\lambda_i \;= \; u_-(p_i)
\; .
\equn
$$

We will also define spinor-strings such as
$$
\BBRQ k {{K}_{i,j}} l \;\equiv\; \BRQ {k^+} {\Slash{K}_{i,j}} {l^+} \;\equiv\; \BRQ
{l^-} {\Slash{K}_{i,j}} {k^-} \;\equiv\; \la l |{{K}_{i,j}}| k] \;\equiv\; \sum_{a=i}^j\spb k.a\spa a.l\,, \equn
$$
as well as
$$
\BRQ k {{K}_{i,j}{K}_{m,n}} l  \;\equiv\; \BRQ {k^-}
{\Slash{K}_{i,j}\Slash{K}_{m,n}}
{l^+}
  \;\equiv\;
\sum_{a=i}^{j}\sum_{b=m}^{n}\spa k.a\spb a.b\spa b.l\,,
\equn
$$
and
$$
 {[k |{{K}_{i,j}{K}_{m,n}} |l]}  \;\equiv\;
{\BRQ {k^+} {\Slash{K}_{i,j}\Slash{K}_{m,n}} {l^-}} \;\equiv\;
\sum_{a=i}^{j}\sum_{b=m}^{n}\spb k.a\spa a.b\spb b.l\,.
\equn
$$
In some cases, we also make use of commutators, such as
$$
\BRQ k {[q, K] } l \;\equiv\; \BRQ k {q K } l \; -\;  \BRQ k {K q} l\,.
\equn
$$


\section{Supersymmetric Decomposition of QCD Amplitudes}

The one-loop amplitudes for gluon scattering
depend on the particle content of the theory since
any particle of the theory with gauge charge may circulate in the loop. Let
$A_{n}^{[J]}$ denote the contribution to gluon scattering due to an
(adjoint representation) particle of spin $J$\,.
The three choices we are primarily
interested in are gluons ($J=1$), adjoint fermions ($J=1/2$) and
adjoint scalars ($J=0$).  Instead of calculating contributions
for these three particle types directly, it is considerably
easier to instead calculate the contributions due to supersymmetric matter
multiplets together with that of a complex scalar (sometimes denoted as
$\NeqZero$ supersymmetry). The three types of supersymmetric multiplet are the
$\NeqFour$ multiplet and the $\NeqOne$ vector and matter
multiplets. These contributions are related to the $A_{n}^{[J]}$
by
$$
\eqalign{
A_{n}^{\,\NeqFour} &\;\equiv\;
A_{n}^{[1]}\; +\; 4A_{n}^{[1/2]}\;+\;3 A_{n}^{[0]}\,,
\cr
A_{n}^{\,\NeqOne\; {\rm vector}} &\;\equiv\; A_{n}^{[1]}\
\;+\;A_{n}^{[1/2]} \,,
\cr
 A_{n}^{\,\NeqOne\; {\rm chiral}} &\;\equiv\;
A_{n}^{[1/2]}\; +\; A_{n}^{[0]} \,.
\cr}
\equn
$$
The contributions from these three multiplets
are not independent but satisfy
$$
A_{n}^{\,\NeqOne\; {\rm vector}} \;\equiv\; A_{n}^{\,\NeqFour} \;-\;
3A_{n}^{\,\NeqOne\; {\rm chiral}} \,.
\equn
$$
We can then invert these
relationships to obtain the amplitudes for QCD via
$$
\eqalign{
A_{n}^{[1]} &\; = \; A_{n}^{\,\NeqFour}-4A_{n}^{\,\NeqOne\; {\rm chiral}}\;+\;A_{n}^{[0]}\,,
\cr
A_{n}^{[1/2]} &\; = \; A_{n}^{\,\NeqOne\; {\rm chiral}}\;-\;A_{n}^{[0]}\,.
\cr}
\equn\label{SusyQCDDecomp}
$$
Although the above relationship is for an adjoint fermion in the loop, the
contribution from a massless fundamental representation quark loop
can be obtained from
the leading colour contribution of the adjoint case~\cite{FiveGluon,BDDKb}.
In this paper, we shall
be focusing on two elements of this supersymmetric decomposition,
namely the $A_{n}^{\,\NeqOne\ {\rm chiral}}$ and $A_{n}^{[0]}$ terms.
Together with the results for $A_{n}^{\,\NeqFour}$ amplitudes these form
the components of QCD amplitudes.

A one-loop
amplitude may be expressed as a linear combination of dimensionally
regularised scalar integral functions $I_i$ with rational (in the
variables $\lambda_a^i$ and $\tilde\lambda_{\dot a}^j$) coefficients
$c_i$\,,~\cite{EarlierIntegrals,Integrals,BDDKa}
$$
A \; = \; \sum_{i}  c_i  I_i \,.
\label{generalform}
\equn$$
In QCD, in general, after series expanding in $\e = (D-4)/2 $, there
are also finite rational function contributions arising from
$\Ord(\e)$ terms striking ultraviolet poles in $\eps$.

The above supersymmetric decomposition~(\ref{SusyQCDDecomp}) of QCD 
amplitudes is useful because
of the differing analytic properties of the components. Each
of the components can be computed separately and then reassembled at
the end to obtain QCD amplitudes~\cite{FiveGluon,BDDKa,BDDKb}.  For
supersymmetric amplitudes the summation is over a smaller set of
integral functions than in the generic non-supersymmetric set.  In
$\NeqFour$ theories, the functions that appear are only scalar box
functions; whereas for $\NeqOne$ theories we are limited to scalar
boxes, scalar triangles and scalar bubbles with no additional rational
pieces. In ref.~\cite{BDDKa,BDDKb} it was demonstrated that
supersymmetric amplitudes are ``cut-constructible'' using only
four-dimensional cuts, {\it i.e.,} the coefficients $c_i$ can be
entirely determined by knowledge of the four-dimensional cuts of the
amplitude.  The scalar loop contributions are, however, not fully
constructible from their four-dimensional cuts and contains additional
rational functions.  Such terms arise, for example, from tensor bubble
integrals.  One may obtain these from cuts as well but the cuts must
be evaluated in dimensional regularisation by working to higher-orders
in the dimensional regularisation
parameter $\epsilon$\,~\cite{BernMorgan,DimShift,QCDUnitarity,TwoLoopSplitting,
Brandhuber2005QCD}.


\section{Recursion Relations for One-Loop Integral Coefficients}
\label{LoopRecursionSection}

In this section we will develop tools for constructing the
coefficients of the integral functions in one-loop amplitudes
recursively.


\subsection{Review of Recursion Relations for Tree Scattering Amplitudes}

We first briefly review the proof~\cite{BCFW} of tree-level on-shell recursion
relations~\cite{TreeRecursion}.  Let us begin by considering a generic
tree amplitude $A(p_1,p_2,\ldots,p_n)$ in complex momentum space and
its behaviour under the following shift of the spinors,
$$
\eqalign{
\tilde \lambda_a \; \rightarrow \;& \tilde \lambda_a \;+\;z\tilde \lambda_{b}\,,
\cr
\lambda_{b} \; \rightarrow \;&  \lambda_{b} \;-\; z \lambda_{a}\,,
\cr}
\equn\label{BasicShift}
$$
where $z$ is an arbitrary complex variable. This will shift the
momentum of the legs $a$ and $b$\,,
$$
\eqalign{
p_a(z) & \; = \; \lambda_a\tilde\lambda_a \;+\; z\lambda_a\tilde
\lambda_{b}\,,\ \ \ \ \cr p_b(z) & \; = \; \lambda_b\tilde\lambda_b \;-\;
z\lambda_a \tilde\lambda_{b}\,.\ \ \ \ \cr}
\equn
$$
By this shift, legs $a$ and $b$ remain on shell, $p^2_{a}(z)=p^2_{b}(z)=0$\,,
and the combination $p_a(z)+p_b(z)$ is independent of the
parameter $z$\,.  Under the shift, the amplitude becomes,
$$
\eqalign{
A(p_1,p_2,\ldots,p_n)
\;\rightarrow\;
A(p_1,p_2,\ldots,p_a(z),\ldots,p_b(z),\ldots,p_n)\;\equiv\; A(z)\,,
\cr}
\equn
$$
where the shift respects the total momentum conservation of the amplitude.
The shifted amplitude $A(z)$ is an analytic continuation of
the physical on-shell unshifted amplitude $A(0)$ into the complex plane.

For a tree amplitude in a gauge theory it has been proven in
ref.~\cite{BCFW} that there will {\it always} be
shifts for which the shifted amplitude $A(z)$ will vanish
as $|z| \rightarrow \infty$\,.  Moreover, it can be shown that
$A(z)$ only has simple poles in $z$ and the only poles which
are present are the same ones which are present for real momenta.
We can then evaluate the following contour integral at infinity,
$$
\frac{1}{2\pi i} \oint {dz\over z}A(z) \; = \; C_\infty
\; = \; A(0) \;+\; \sum_\alpha  {\rm Res}_{z=z_\alpha}{A(z) \over z}\,,
\equn
$$
where $z_\alpha$ are the simple poles of $A(z)$ in $z$ and ${\rm
Res}_{z=z_\alpha}$ signifies the residue at the location of the poles.
Using the vanishing of $A(z)$ as $|z| \rightarrow \infty$
the boundary term $C_\infty$ also vanishes and
$$
A(0) \; = \; -\sum_\alpha  {\rm Res}_{z=z_\alpha}{A(z) \over z}\,.
\equn
$$

The poles and residues in the shifted amplitude $A(z)$ are determined
by the factorisation properties of the on-shell amplitude. An on-shell
amplitude will factorise into a product of two on-shell tree
amplitudes as $K_{i,j}^2\equiv (k_i+\cdots+k_j)^2 \rightarrow 0$\,, {\it
i.e.} we will have the following factorisation
$$
A\;\inlimit^{K_{i,j}^2\rightarrow 0}\;
A( k_i, \cdots,  k_j, K_{i,j}) \times {i \over K_{i,j}^2 }\times
A( k_{j+1}, \cdots, k_{i-1} ,-K_{i,j})\,.
\equn
$$
At tree-level one can show that there are no other factorisations
in the complex plane.  Using these factorisation limits one can then
write the following recursion relation for tree amplitudes
$$
A(0) \; = \;  \sum_{\alpha,h}
 {A^h_{n-m_\alpha+1}(z_\alpha)
{i\over K^2_{\alpha}}A^{-h}_{m_\alpha+1}(z_\alpha)}\,,
\equn\label{RecursionTree}
$$
where $A^h_{n-m_\alpha+1}(z_\alpha)$ and
$A^{-h}_{m_\alpha+1}(z_\alpha)$ are shifted
amplitudes evaluated at the residue value $z_\alpha$\,, and $h$ denotes
the helicity of the intermediate state corresponding to the propagator
term $i/K^2_{\alpha}$\,.  In eq.~(\ref{RecursionTree}) the sum selects
tree amplitudes where the reference leg $p_a$ is contained in one tree
amplitude and the reference leg $p_b$ is contained in the other.


\subsection{Recursion Relations for Integral Coefficients}\label{rerelfintco}
\label{IntegralCoeffRecursionSubsection}

For the special set of one-loop amplitudes which consist of only
rational functions (the ``finite'' loop amplitudes), a set of
recursion relations were developed in
ref.~\cite{BDKrecursionA,BDKrecursionB} for determining all such
amplitudes.  One-loop recursion relations
amplitudes are more subtle than at tree-level because of the
appearance of double and `unreal' poles whose nature are not yet fully
understood.  More generally the presence of branch cuts invalidates a
basic assumption of the tree-level proof outlined above.

Very recently, a unitarity-factorisation bootstrap method was derived
for dealing with the more general case containing cuts~\cite{BDKSix}, systematising an
earlier unitarity-factorisation bootstrap
approach~\cite{ZFourPartons}.  With this method the cut containing
terms are evaluated using the unitarity method and the rational
function pieces by finding an appropriate on-shell recursion relation,
accounting for overlaps between the two types of terms.  In practice
many results are already known for the cut containing pieces, so in
these cases only the rational function parts need to be determined in
order to have complete QCD amplitudes.  As an example, an
explicit construction of the rational parts of the six-gluon QCD
amplitude $A_{n;1}^{\rm QCD}(1^-, 2^-, 3^+, 4^+, 5^+, 6^+)$ was
given in ref.~\cite{BDKSix}.

Here we wish to address the complementary question of whether we can
also apply recursion relations to the cut containing pieces.  The
approach we take will be to construct recursion relations, not
for amplitudes but for the coefficients of the integrals since these
are cut free rational functions.  In this way we avoid the issue of
constructing recursion relations in the presence of branch cuts.
However, to be successful, a firm grasp of the analytic properties of
the integral coefficients is required. These properties differ from
those of full amplitudes: in general these coefficients will not
contain the complete set of poles appearing in the amplitudes and they
will contain spurious poles.  The spurious poles can interfere with the
construction of recursion relations since there are no theorems to
guide factorisations on spurious denominators. These spurious poles in
general cancel only in the sum over all terms in the amplitude, and
not in individual integral coefficients.  But with an understanding of
which spurious poles appear in which integral coefficients, we can
maneuver around them.  A detailed discussion of the factorisation
properties of one-loop amplitudes, as well the spurious singularities
that appear, may be found in refs.~\cite{BDDKa,BernChalmers}.  Here we will
only summarise a few salient points.

First consider the factorisation where a three vertex is isolated as
one of the factors.  For massless particles and real momenta such
vertices vanish on-shell, though not for complex
momenta~\cite{TreeRecursion}. The real momentum factorisation is more
conventionally presented in terms of limits where the momenta of two
legs $p_a$ and $p_b$ become
collinear~\cite{BDDKa,BDDKb,BernChalmers,KilgoreSplit,KosowerSplit},
\vspace{-0.0cm}
$$\hspace{0.4cm}
\eqalign{
A^{\oneloop}_n(\ldots,p_a,p_b,\ldots)\ \  \inlimit^{a \parallel b} \ \ &
\sum_{h}
{\rm Split}^\tree_{-h}(a^{h_a},b^{h_b})
A^{\oneloop}_{n-1}(\ldots,(P)^h,\ldots)
\cr  + &\sum_{h}
{\rm Split}^{\oneloop}_{-h}(a^{h_a},b^{h_b})
A^\tree_{n-1}(\ldots,(P)^h,\ldots)\,,}
\equn
$$
where $P = p_a + p_b$\,.  The case of multi-particle factorisation is
also well understood; one-loop amplitudes factorise according to
a universal formula~\cite{BernChalmers},
$$
\hspace{-0.2cm}\eqalign{
A_{n}^{\oneloop}\
& \hskip -.12 cm
\ \inlimit^{K_{_{i,i+m-1}}^2 \hspace{-0.4cm}\longrightarrow\,\, 0}\
\hskip .15 cm
\sum_{h=\pm}  \Biggl[
 A_{m+1}^{\oneloop}(\ldots ,K_{i,i+m-1}^h\,,\ldots) \,
            {i \over K_{i,i+m-1}^2} \,
   A_{n-m+1}^{\tree}(\ldots ,(-K_{i,i+m-1})^{-h}\,, \ldots) \cr
& \hskip2.8cm \null
 + A_{m+1}^{\tree}(\ldots, K_{i,i+m-1}^h\,,\ldots) \, {i\over K_{i,i+m-1}^2} \,
   A_{n-m+1}^{\oneloop}(\ldots,(-K_{i,i+m-1})^{-h}\,,\ldots)
\label{LoopFact} \cr
& \hskip-1.5cm \null
 + A_{m+1}^{\tree}(\ldots, K_{i,i+m-1}^h\,,\ldots) \, {i\over K_{i,i+m-1}^2} \,
   A_{n-m+1}^{\tree}(\ldots,(-K_{i,i+m-1})^{-h}\,, \ldots) \,
      \Fact_n(K_{i,i+m-1}^2;p_1, \ldots, p_n) \Biggr] \,,
\cr}\equn
\label{factorisation}
$$
where $K_{i,i+1-1}^2$ is the momentum invariant on which the amplitude
factorises. The `factorisation function' denoted by
$\Fact_n(K_{i,i+m-1}^2;p_1, \ldots, p_n)$ actually represents a
non-factorisation, since it contains kinematic invariants which cross
the pole; they  however have a universal structure linked to
infrared divergences and are given explicitly
in ref.~\cite{BernChalmers}.

The factorisation of amplitudes follows from
the combined behaviour of integral functions and the integral
coefficients in the factorisation limit.  If we turn this around,
given the general factorisation (\ref{factorisation}) of an amplitude
and given its basis of integral functions, we may then determine
the factorisation properties of the integral coefficients.

Consider first the collinear limits.  If we expand the amplitude
in terms of integrals times coefficients using \eqn{generalform}, we
have that
$$\hspace{-0.7cm}
\eqalign{
\sum_i c_{i,n}  I_{i,n}\ \  \inlimit^{a \parallel b} \ \ &  \sum_{h}
{\rm Split}_{-h}^\tree(a^{h_a},b^{h_b})
\sum_i c_{i,n-1}^h  I_{i,n-1}
  + \sum_{h}
{\rm Split}_{-h}^{\oneloop}(a^{h_a},b^{h_b}) A^{h\,\tree}_{n-1}\,.}
\equn\label{SplitCoeffs}
$$
In order to disentangle the behaviour of the coefficients we need
to know how the integrals flow into each other under collinear
factorisation~\cite{BDDKa,BernChalmers}.
It is useful to choose a good basis of integral functions with
appropriate factors moved between coefficients and integrals
in order to have a simple factorisation behaviour.  We will
make one such choice
below when we discuss the examples.

Now focus on the behaviour of a single term $c_{i,n}I_{i,n}$ in the
sum (\ref{SplitCoeffs}) as two momenta become collinear.  The very
simplest case is when the two momenta $p_a$ and $p_b$ becoming
collinear belong to legs in a single cluster of external legs of the
integral.  In the case that the cluster has at least three legs so it
does not become massless, the integrals have a smooth degeneration in
the collinear limits as illustrated,

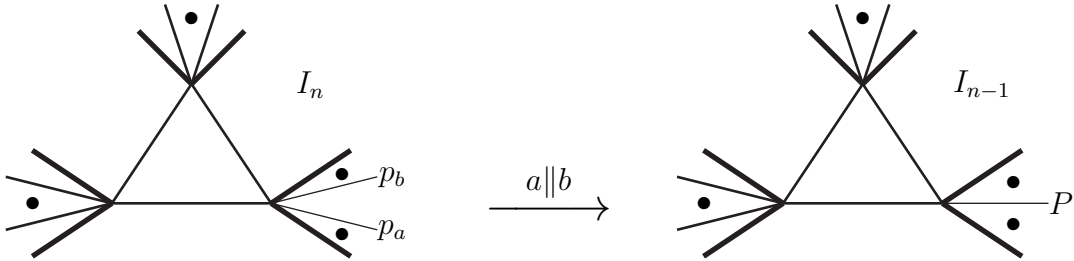
\vspace{-0.4cm}
\begin{center}\hspace{0.4cm}
\begin{picture}(100,70)(40,60)
\Line(30,30)(70,40)
\Line(30,30)(70,20)
\SetWidth{2}
\Line(30,30)(60,50)
\Line(30,30)(60,10)
\SetWidth{1}
\Line(30,30)(-30,30)
\Line(-30,30)(0,75)
\Line(30,30)(0,75)
\Text(76,20)[]{$p_a$}
\Text(76,40)[]{$p_b$}
\Text(135,35)[]{\Large$\inlimit^{a \parallel b}$}
\Line(-30,30)(-70,40)
\Line(-30,30)(-70,20)
\SetWidth{2}
\Line(-30,30)(-60,50)
\Line(-30,30)(-60,10)
\SetWidth{1}
\Text(-60,30)[]{$\bullet$}
\Line(0,75)(-10,105)
\Line(0,75)(10,105)
\SetWidth{2}
\Line(0,75)(-20,95)
\Line(0,75)(20,95)
\Text(0,100)[]{$\bullet$}
\Text(45,75)[]{$I_{n}$}
\Text(57,41)[]{$\bullet$}
\Text(57,18)[]{$\bullet$}

\end{picture}
\begin{picture}(100,70)(-110,60)
\Line(30,30)(70,30)
\SetWidth{2}
\Line(30,30)(60,10)
\Line(30,30)(60,50)
\SetWidth{1}
\Line(-30,30)(0,75)
\Line(30,30)(-30,30)
\Line(30,30)(0,75)
\Text(45,75)[]{$I_{n-1}$}
\Text(75,30)[]{$P$}
\Line(-30,30)(-70,40)
\Line(-30,30)(-70,20)
\SetWidth{2}
\Line(-30,30)(-60,50)
\Line(-30,30)(-60,10)
\SetWidth{1}
\Text(-60,30)[]{$\bullet$}
\Line(0,75)(-10,105)
\Line(0,75)(10,105)
\SetWidth{2}
\Line(0,75)(-20,95)
\Line(0,75)(20,95)
\SetWidth{1}
\Text(0,100)[]{$\bullet$}
\Text(57,38)[]{$\bullet$}
\Text(57,22)[]{$\bullet$}
\end{picture}

\end{center}
\vspace{1.7cm}
where $P=p_a + p_b$ becomes massless in the factorisation limit.

\noindent
In this case the factorisation of the coefficient is simply
$$
c_{i,n} \ \inlimit^{a \parallel b} \ \
\sum_h {\rm Split}_{-h}^\tree(a^{h_a},b^{h_b})\;  c_{i,n-1}^h\,,
\equn
$$
where the coefficient on the left belongs to the left integral
in the figure and the coefficient on the right belongs to the
right integral.   The coefficient has this simple behaviour
because in this case there are no contributions to the
${\rm Split}_{-h}^{\oneloop}$ term in \eqn{SplitCoeffs}.  Contributions
from this term all come either from discontinuities as massive
clusters degenerate to massless ones
or from integrals with two massless legs carrying the collinear
momenta $p_a$ and $p_b$\,. (See ref.~\cite{BernChalmers} for further details.)

By applying the same logic to multi-particle factorisations
(\ref{factorisation}) we conclude that the coefficients
also behave as if they were tree amplitudes
as long as the factorisations are entirely within
a cluster of legs (and are not on the momentum invariant of
the entire cluster).  That is the coefficient behaves as,
$$
c_{i,n} \ \inlimit^{K^2 \rightarrow 0} \ \
\sum_h   A^h_{n-m+1}\,\frac{i}{K^2}\,c_{i,m+1}^{-h}\,.
\equn
$$

Assuming that the spurious denominators do not pick up
a $z$ dependence --- below
we describe simple criteria for ensuring this  ---
we  obtain a recursion relation for the coefficients
which strikingly is no more complicated than for tree amplitudes,
$$
c_n(0) \; = \; \sum_{\alpha,h}  {A^h_{n-m_\alpha+1}(z_\alpha) \,
 {i\over K^2_{\alpha}}\, c^{-h}_{m_\alpha+1}(z_\alpha)} \,,
\equn\label{CoeffRecur}
$$
where $A^h_{n-m_\alpha +1}(z_\alpha)$ and $c^h_{n-m_\alpha+1}(z_\alpha)$ are
shifted tree amplitudes and coefficients evaluated at the residue
value $z_\alpha$\,, $h$ denotes the helicity of the intermediate state
corresponding to the propagator term $i/K^2_{\alpha}$\,. In this
expression one should only sum over a limited set of poles; if the
shifts are chosen from within a cluster, the only poles that should
be included are from within the kinematic invariants formed from the
momenta making up the cluster. Pictorially, this coefficient
recursion relation is,
\vspace{-2.8cm}
\begin{center}
\begin{picture}(70,70)(20,60)
\Line(30,30)(80,30)
\Line(30,30)(10,40)
\Line(30,30)(10,20)
\Line(30,30)(50,40)
\Line(30,30)(50,20)
\Line(30,30)(20,50)
\Line(30,30)(20,10)
\Line(80,30)(100,40)
\Line(80,30)(100,20)
\BCirc(30,30){10} \BCirc(80,30){10} \Text(80,30)[]{$A_\alpha$}
\Text(30,30)[]{$c_\alpha$}
\Text(-20,30)[]{\huge $\sum$}
\Text(-21,11)[]{ $\alpha$}
\Text(15,43)[]{ $\bullet$}
\Text(15,17)[]{ $\bullet$}
\Text(95,30)[]{$\bullet$}
\end{picture}
\end{center}
\vspace{1.5cm}
\label{recursion}

In order to have a valid bootstrap we must have that the shifted
coefficient vanishes as $|z|\rightarrow \infty$; otherwise there would
be a dropped boundary term.  We can, however, impose criteria to
prevent this from happening.  Consider an integral and consider the
unitarity cut which isolates the cluster on which the recursion will
be performed, {\it i.e.} the one with the two shifted legs,

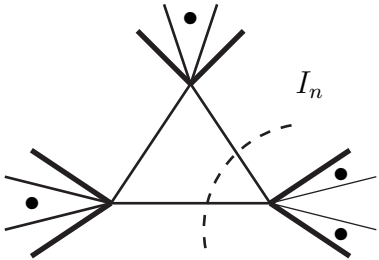
\vspace{-0.3cm}
\begin{center}
\begin{picture}(40,60)(-20,50)
\Line(30,30)(70,40)
\Line(30,30)(70,20)
\SetWidth{2}
\Line(30,30)(60,50)
\Line(30,30)(60,10)
\SetWidth{1}
\Line(30,30)(-30,30)
\Line(-30,30)(0,75)
\Line(30,30)(0,75)
\Line(-30,30)(-70,40)
\Line(-30,30)(-70,20)
\SetWidth{2}
\Line(-30,30)(-60,50)
\Line(-30,30)(-60,10)
\SetWidth{1}
\Text(-60,30)[]{$\bullet$}
\Line(0,75)(-10,105)
\Line(0,75)(10,105)
\SetWidth{2}
\Line(0,75)(-20,95)
\Line(0,75)(20,95)
\Text(0,100)[]{$\bullet$}
\Text(45,75)[]{$I_{n}$}
\Text(57,41)[]{$\bullet$}
\Text(57,18)[]{$\bullet$}
 \SetWidth{1}
\DashCArc(45,20)(40,100,190){4}
\end{picture}
\end{center}
\vspace{1.2cm}
\noindent
The dashed line in this figure indicates the cut. The recursion is to
be performed with the two shifted legs from the right-most
cluster. Then simple criteria for a valid recursion are:
\begin{enumerate}

\item The shifted tree amplitude, on the side of the cluster
undergoing recursion, vanishes as $|z| \rightarrow \infty$\,.

\item All loop momentum dependent kinematic poles
are unmodified by the shift (\ref{BasicShift}).

\end{enumerate}
Since the location of none of the poles in $z$ depends on loop
momentum and any shifted tree amplitude can be written in the form of
poles in $z$ times residues (which are of course independent of $z$),
the $z$ dependence appears only as an overall prefactor in front of the
loop integral.  We are therefore guaranteed that the shifted
coefficient will have vanishing behaviour as $|z|\rightarrow \infty$\,.
With these two criteria we have an added bonus: The spurious
singularities appearing in the shifted coefficients cannot depend on
$z$ since the integration does not tangle with $z$\,.  This guarantees
that our recursion relations are not contaminated by spurious
singularities.  In a direct integration of the cuts, spurious
singularities would arise from evaluating the integrals, but since
$z$-dependence does not enter into the integrals they cannot enter
into spurious singularities.  All calculations that we perform in this
paper satisfy these two conditions proving that there are no
boundary terms or spurious singularities invalidating the recursion.

The above criteria are simple sufficiency conditions for constructing a
recursion for loop integrals, but are not necessarily required:
It is possible to find valid shifts outside the above class, for
example, involving shifted legs from different corners of triangle or
box integrals, but we will not discuss this here.  In general, one
would also need to account for factorisations involving the second
term in \eqn{SplitCoeffs} as well as non-trivial flows under
factorisation of integral functions into each other and into loop
splitting and factorisation functions~\cite{BernChalmers}.  In
addition some care is required to avoid shifting the spurious
denominators.  We leave a discussion of these issues to future studies.

\subsection{Example}

To make our procedure more concrete, it is useful to illustrate it with an
explicit example, before turning to the main calculation of this
paper.  In this section we will use recursion relations to compute the
coefficients of the six-point split-helicity one-loop amplitudes in the
$\NeqOne$ chiral multiplet as well as for the scalar loop contribution
$A_{n;1}^{[0]}$\,.

In general, one-loop amplitudes contain a set of integral
functions. For the amplitudes of interest here, a convenient set
is~\cite{FiveGluon},
$$
\eqalign{
\Kz(r) & \; = \; {1 \over \epsilon\,{(1-2\,\epsilon)}}(-r)^{-\epsilon}
\; = \; \Big(-\log(-r)\;+\;2\;+\;{1\over \epsilon}\Big)
\;+\; {\cal{O}}(\epsilon)\,,\cr
\Lz(r) & \; = \; {{\log(r) \over 1\;-\;r}}\,, \hspace{2.3cm}
\Lzz(r) \; = \; {{\log(r) \;-\; {(r\;-\; 1/r)/2}} \over {(1\;-\;r)^3}}\,.
}\equn
$$
We find it convenient to use the $\Lz$\,, $\Lzz$ and $\Kz$ functions
to build up the expressions for the complete amplitudes.  This choice
of integral functions is a good one from the viewpoint of having
simple factorisation properties for the integral coefficients.  The
$\Lz$ and $\Lzz$ functions can be thought of either as the difference
of two bubble integrals or as a two-mass triangle with an integrand
which is linear in the Feynman parameter for the leg linking the two
massive legs.  The $\Lzi$ functions have the advantage of being finite
in $\eps$ and free from spurious singularities as $r\rightarrow 1$\,.

The dimensionally regularised amplitudes carry an overall factor of
$c_\Gamma$\,,
$$
c_\Gamma\; = \;
{1\over(4 \pi)^{2\,-\,\eps}}\,{\Gamma(1\,+\,\eps)\,\Gamma^2(1\,-\,\eps)\over\Gamma(1\,-\,2\,\eps)}\,,
\equn
$$
which we suppress in this paper.

In our example, we will use a five-gluon one-loop amplitude
with an $\NeqOne$ chiral multiplet in the loop as the starting point
in the recursion. This five-gluon amplitude is~\cite{FiveGluon},
$$
\hspace{0.3cm}
\eqalign{
A^{\,\NeqOne\,\,{\rm chiral}}(1^-,2^-,3^-,4^+,5^+) &
\; = \; \frac{1}{2}\,A^{\tree}(1^-,2^-,3^-,4^+,5^+)\,
\big(\Kz(t_{5,1})+\Kz(t_{3,4})\big)\cr &\;+\frac{1}{2}\,
c^{\,\NeqOne\,\,{\rm chiral}}(5^+,1^-;2^-;3^-,4^+)
\frac{\Lz\big(-t_{5,1}/(-t_{5,2})\big)}{t_{5,2}}\,,}
\equn
$$
where,
$$
\eqalign{
A^{\tree}(1^-,2^-,3^-,4^+,5^+)&
\;\equiv\;-i\,\frac{[4\,5]^3}{[5\,1][1\,2][2\,3][3\,4]}\,,\cr
c^{\,\NeqOne\,\,{\rm chiral}}(5^+,1^-;2^-;3^-,4^+)&
\;\equiv\;-i\,\frac{[4\,5]^2\,\left[4|[k_2,K_{5,1}]|5\right]}
{[5\,1][1\,2][2\,3][3\,4]}\,.
}\equn
$$
In general we will use semicolons to delineate the groupings of
the legs into clusters of the associated triangle integrals. In this
case the triangle diagram representation of $\Lz(-t_{5,1}/(-t_{5,2}))$ is
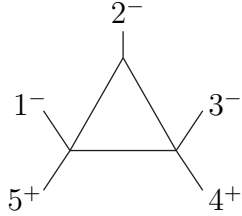
\begin{center}\vspace{-.5 cm}
\begin{picture}(100,100)(0,0)
\Line(30,30)(70,30)
\Line(30,30)(50,65)
\Line(50,65)(70,30)

\Line(30,30)(20,45)
\Line(30,30)(20,15)

\Line(70,30)(80,45)
\Line(70,30)(80,15)

\Line(50,65)(50,75)

\Text(52,83)[]{$2^-$}

\Text(15,48)[]{$1^-$}
\Text(13,12)[]{${5}^+$}

\Text(89,48)[]{$3^-$}
\Text(89,12)[]{$4^+$}

\Text(-20,55)[]{$$}
\Text(50,43)[]{$$}
\end{picture}
\end{center}
\vspace{-.4cm}

For the  cut-containing terms of the
scalar loop contribution to the five-gluon amplitude
$$
\eqalign{
A^{[0]}(1^-,2^-,3^-,4^+,5^+)&\; = \;
\frac{1}{3}A^{\,\NeqOne\,\,{\rm chiral}}(1^-,2^-,3^-,4^+,5^+)\cr
&\;+\frac{1}{3}\,c^{[0]}(5^+,1^-;2^-;3^-,4^+)\,
\frac{\Lzz\big(-t_{5,1}/(-t_{5,2})\big)}{(t_{5,2})^3}
+ \hbox{ rational}\,,
}
\equn
$$
where the coefficient $c^{[0]}$ is given by,
$$
c^{[0]}(5^+,1^-;2^-;3^-,4^+)\; = \; -i\,
\frac{[4|k_2K_{5,1}|5]\,[4|K_{5,1}k_2|5]\,
\left[4|[k_2,K_{5,1}]|5\right]\,}{[5\,1][1\,2][2\,3][3\,4]} \,.
\equn
$$
In this case the rational terms in the amplitude are
known~\cite{FiveGluon}, but we will not need these here. (A very
recent discussion of rational parts is given in
ref.~\cite{BDKSix}.)  Note that the two coefficients
$c^{\,\NeqOne\,\,{\rm chiral}}$ and $c^{[0]}$ have different
dimensions, because they multiply factors of different dimensions.
(The functions $\Lz$ and $\Lzz$ are dimensionless.)

The example here illustrates how recursion relations for loop
coefficients work and will lay the ground for the generic computation
of the $n$-point split-helicity coefficients
$c_n^{\,\NeqOne\,\,{\rm chiral}}$ and $c_n^{[0]}$\,.  As we will see,
the coefficients $c_n^{\,\NeqOne\,\,{\rm chiral}}$ and $c^{[0]}_n$ are
computed as sums over different shifts of the five-point coefficients
$c_5^{\,\NeqOne\,\,{\rm chiral}}$ and $c_5^{[0]}$\,. It will turn out
that these coefficients are quite similar and we can therefore unify
the expression for the coefficients using a variable $\Slash{\!P}$ and
$\Slash{\!\tilde P}$ which both takes the identity value $\mathbb{I}$
for the $\NeqOne$ chiral coefficients, where $\mathbb{I}$ is defined to
be the identity matrix in the space of gamma matrices and
$\Slash {\!P}=\Slash{k_2}\Slash{\!K_{5,1}}$\,,
$\Slash {\!\tilde P}=\Slash {\!K_{5,1}}\Slash {k_2}$ in
the scalar loop contribution.  The five-point `googly' coefficients
$c_5$ for both theories can then be written as,
$$
\eqalign{
c_5&=c(5^+,1^-;2^-;3^-,4^+)\;=\;-i
\frac{[4|P|5]\,[4|\tilde P|5]\,\left[4|[k_2,K_{5,1}]|5\right]\,
}{[5\,1][1\,2][2\,3][3\,4]}\,.}
\equn\label{eqfivepointMHV}
$$
(In twistor terminology, `googly' means the conjugate amplitude
where positive and negative helicities are interchanged. In the MHV
case, a googly amplitude has two positive and the rest negative
helicities.)

Now consider the six-point coefficient
$c_6=c_6(6^+,1^-;2^-;3^-,4^+,5^+)$
of the integral functions $\Lz(-t_{6,1}/(-t_{6,2}))$ and
$\Lzz(-t_{6,1}/(-t_{6,2}))$\,. Graphically the associated
integral function is,
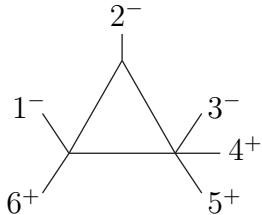
\begin{center}\vspace{-0.8cm}
\begin{picture}(100,100)(0,0)
\Line(30,30)(70,30)
\Line(30,30)(50,65)
\Line(50,65)(70,30)

\Line(30,30)(20,45)
\Line(30,30)(20,15)

\Line(70,30)(80,45)
\Line(70,30)(87,30)
\Line(70,30)(80,15)

\Line(50,65)(50,75)

\Text(52,83)[]{$2^-$}

\Text(15,48)[]{$1^-$}
\Text(13,12)[]{${6}^+$}

\Text(89,48)[]{$3^-$}
\Text(97,32)[]{$4^+$}
\Text(89,12)[]{$5^+$}
\end{picture}
\end{center}\vspace{-0.6cm}
\noindent
In this example we will use the shift,
$$\hspace{1cm}\eqalign{
  \tilde\lambda_3&\;\rightarrow\;\tilde\lambda_3\;-\;z\tilde\lambda_4\,,\cr
  \lambda_4&\;\rightarrow\;\lambda_4\;+\;z\lambda_3\,.}
\equn\label{Shift34}
$$
\noindent
We will first check that the two criteria for finding valid shifts
are satisfied. Since the shifted legs are from the  $\{3,4,5\}$ cluster, we
must check the cut
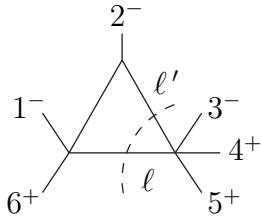
\begin{center}\vspace{-0.8cm}
\begin{picture}(100,100)(0,0)
\Line(30,30)(70,30)
\Line(30,30)(50,65)
\Line(50,65)(70,30)

\Line(30,30)(20,45)
\Line(30,30)(20,15)

\Line(70,30)(80,45)
\Line(70,30)(87,30)
\Line(70,30)(80,15)

\Line(50,65)(50,75)

\Text(52,83)[]{$2^-$}

\Text(15,48)[]{$1^-$}
\Text(13,12)[]{${6}^+$}

\Text(89,48)[]{$3^-$}
\Text(97,32)[]{$4^+$}
\Text(89,12)[]{$5^+$}

\DashCArc(80,20)(30,110,190){4}
\Text(67,57)[]{$\ell\,'$}
\Text(60,20)[]{$\ell$}
\end{picture}
\end{center}\vspace{-0.6cm}
%
For the case of a scalar circulating in the loop the tree amplitude
isolated on the right side of the cut is,
$$
A^\tree(3^-, 4^+, 5^+, \ell_{s}^+, {\ell\,'}\!\!_s\,^-)\; =\;
 i  {\spa3.{\ell_s}^2 \spa3.{\ell\,'\!\!_s}^2 \over \spa3.4 \spa4.5 \spa5.{\ell_s}
       \spa{\ell_s}.{{\ell\,'}\!\!_s} \spa{{\ell\,'}\!\!_s}.3 } \,,
\equn
$$
where the subscript $s$ denotes a scalar circulating in the loop.
Under the shift (\ref{Shift34}), the tree amplitude becomes,
$$
A^\tree(3^-, 4^+, 5^+, \ell_{s}^+, {\ell\,'}\!\!_{s}\,^-; z)\; =\;
 i  {\spa3.{\ell_s}^2 \spa3.{{\ell\,'}\!\!_s}^2 \over \spa3.4 (\spa4.5 + z \spa3.5)
           \spa5.{\ell_s}
       \spa{\ell_s}.{{\ell\,'}\!\!_s} \spa{{\ell\,'}\!\!_s}.3 } \,,
\equn\label{Tree345}
$$
and we immediately see that both criteria are satisfied: the
$z$-dependence factors out of the integrand and the shifted
coefficient times the integral vanishes as $|z| \rightarrow \infty$\,,
because an overall prefactor vanishes.  Similarly, it is easy
to check that the criteria are satisfied also for a fermion
circulating in the loop.  (The $\NeqOne$ chiral case is just the sum
of the fermion and scalar loop contributions so it too satisfies the criteria.)

From \eqn{Tree345} it is clear that the only channel which can
contribute to the recursion is the $\spa4.5$ channel,
which is determined by the collinear factorisation,
$$
\hspace{1cm}\eqalign{
c_6(6^+,1^-;2^-;3^-,4^+,5^+)\ \inlimit^{4 \parallel 5}\
 {\rm Split}^\tree_-(4^+,5^+)  \,c_5(6^+,1^-;2^-;3^-,(4+5)^+)\,.}
\equn
$$
Thus the coefficient $c_6(z)=c(p_6,p_1;p_2;p_3(z),p_4(z),p_5)$ has only
one pole in $z$\,. Notice that the coefficient
$c_5(6^+,1^-;2^-;3^-,(4+5)^+)$ is given by the five-point
amplitude (\ref{eqfivepointMHV}).

Following the tree-level construction described in ref.~\cite{TreeRecursion},
the coefficient $c_6$ can hence be computed as the residue of
$-c_6(z)/z$ at the collinear pole $z_1=-\la 4\,5\ra/\la3\,5\ra$\,,
$$\hspace{1cm}\eqalign{
c_6=c_6(0)=-{\rm Res}_{z=z_1}\frac{c_6(z)}{z}\,.}
\equn
$$
Applying the recursion (\ref{CoeffRecur}),
$$\hspace{1.5cm}\eqalign{
z_1&\;=\;-\la 4\,5\ra/\la3\,5\ra\,,\hspace{4.75cm}\quad \omega\bar\omega\;=\;\la3|K_{4,5}|4]\,,\cr
[\hat 4\,\hat K_{4,5}]&\;=\;[4|K_{4,5}|3\ra/\bar\omega\,,\hspace{4.05cm}\quad [5\,\hat K_{4,5}]\;=\;[5|K_{4,5}|3\ra/\bar\omega\,,\cr
[2\,\hat 3]&\;=\;[2\,3]-z[2\,4]\;=\;[2|K_{3,4}|5\ra/\la3\,5\ra\,,\hspace{0.55cm}\quad [\hat3\,\hat K_{4,5}]\;=\;t_{3,5}/\bar\omega\,,\cr
}
\equn
$$
the triangle coefficient evaluates to,
$$\hspace{1.5cm}\eqalign{
c_6&\;=\;c(6^+,1^-;2^-;\hat 3^-,\hat K_{45}^+)\,\frac{i}{s_{45}}\,A(\hat 4^+,5^+,(-\hat K_{45})^-)\,,\cr
&\;=\; -i\,\frac{[\hat K_{45}|P|6]\,[\hat K_{45}|\tilde P|6]\,\left[\hat K_{45}|[k_2,K_{6,2}]|6\right]}{[6\,1][1\,2][2\,\hat 3]
[\hat 3\,\hat K_{45}]}\,\,\frac{i}{s_{45}}\,\,\frac{(-i)\,[\hat 4\,5]^3}
{[(-\hat K_{45})\,\hat 4][5\,(-\hat K_{45})]}\,,\cr
&=i\,\,\frac{\la 3|K_{3,5}P|6]\,\la 3|K_{3,5}\tilde P|6]\,\la 3|K_{3,5}[k_2,K_{6,2}]|6]}{\BBRQ 2 {K_{3,5}}5\spb6.1\spb1.2\spa3.4\spa4.5\, t_{3,5}}\,.
}
\equn\label{C6p}
$$
The
above expression for the six-point coefficient for $\NeqOne$ agrees
with the result first given in ref.~\cite{BBBDDNEq1}, after setting
$\Slash{\! P}$ and $\Slash{\!{\tilde{P}}}$ equal to unit matrices.
In the scalar loop case, $\Slash {\!P} \equiv\Slash{k_2}\Slash{\!K_{6,1}}$ and
$\Slash {\!\tilde P} \equiv\Slash {\!K_{6,1}}\Slash {k_2}$ in \eqn{C6p}.


\section{All-$n$ Split-Helicity Integral Coefficients}

In this section,
we
present the coefficients of all integral functions in
the split-helicity amplitudes,
$$
A(1^-,2^-,\cdots, l^-,l+1^+,\cdots  n^+ )\,, \equn
$$ with either an $\NeqOne$ super-multiplet or scalar circulating in
the loop.  Within the supersymmetric decomposition of QCD
amplitudes~(\ref{SusyQCDDecomp}), the $\NeqFour$ component always
consists entirely of box integrals~\cite{BDDKa}.  For the split helicity
configuration it is not difficult to show that for an $\NeqOne$ chiral
multiplet or a scalar in the loop there are no boxes and the
amplitudes consist entirely of triangle and bubble integrals~\cite{BBBDDNEq1}.

The starting expressions in our recursion are the one-loop five-point
split-MHV amplitudes which were first obtained in
ref.~\cite{FiveGluon}.
As before,  we use the $\Lzi$ and $\Kz$ as
the basis functions.  From an examination of the unitarity cuts it is
not difficult to show that the only logarithms that appear in the amplitude, to finite order in $\epsilon$,
have arguments $t_{i,j}$ where $1\leq i \leq l$ and $l+1\leq j\leq n$:
otherwise a vanishing tree amplitude appears in the cuts.

Before continuing, it is convenient to employ a unified notation for
both the $\NeqOne$ and scalar loop cases, so that we can present the
computations in the two cases simultaneously.  We define,
$$
\eqalign{
\Slash{\! P} & \;\equiv\; \mathbb{I}
 \,, \hskip 2.2 cm
\Slash{\!{\tilde{P}}} \;\equiv\;  \mathbb{I} \,, \hskip 2.4cm
        \hbox{for the chiral $\NeqOne$ loop contribution\,,} \cr
\Slash{\! P} & \;\equiv\; \Slash{k_m} \Slash {\!K_{L}}
 \,, \hskip 1.3 cm
\Slash{\!{\tilde{P}}} \;\equiv\; \Slash {\!K_{L}}\Slash {k_m}\,, \hskip 1.47cm
        \hbox{for the scalar loop contribution\,,}
}
\equn\label{PDef}
$$
where $\mathbb{I}$ is the identity matrix, $k_m$ is the momentum
of the massless corner,  and
$K_L$ denotes the total momentum of the left massive corner of the triangle,
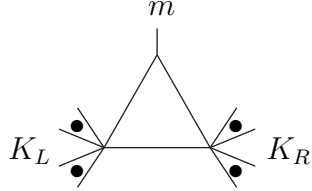
\vspace{-0.5cm}
\begin{center}
\begin{picture}(100,100)(0,0)
\Line(30,30)(70,30) \Line(30,30)(50,65) \Line(50,65)(70,30)
\Line(30,30)(20,45) \Line(30,30)(13,23) \Line(30,30)(13,37) \Line(30,30)(20,15)
\Line(70,30)(80,45) \Line(70,30)(87,37) \Line(70,30)(87,23) \Line(70,30)(80,15)
\Line(50,65)(50,75)
\Text(52,83)[]{$m$}

\Text(20,38)[]{\textcolor{red}{$\bullet$}}
\Text(20,21)[]{\textcolor{blue}{$\bullet$}}
\Text(2,29)[]{$K_L$}

\Text(80,38)[]{\textcolor{red}{$\bullet$}}
\Text(100,29)[]{$K_R$}
\Text(80,21)[]{\textcolor{blue}{$\bullet$}}

\Text(-20,55)[]{$$} \Text(50,43)[]{$$}
\end{picture}
\end{center}
\vspace{-0.6cm}

We write the amplitude for an $\NeqOne$ chiral multiplet in a $\Lz$ and
$\Kz$ basis so that,
$$\hspace{0.1cm}
\eqalign{
A^{\,\NeqOne\ {\rm chiral}}(1^-,2^-,\cdots  l^-,l+1^+,\cdots, n^+) \;=\; &\,\,
 a_1\, \Kz( s_{n1} )\;+\;a_2\, \Kz( s_{ll+1} )\cr
& \;\hspace{-5.92cm}\;+\; \frac{1}{2}\sum_{m,r}^{}\,\, c^{m,r}_{n,l}\,\,
         { {\rm L}_{0} [ t_{m,r}
 / t_{m+1,r} ] \over t_{m+1,r}  }\; +\; \frac{1}{2}\sum_{ \bar m, \bar r}\,\, \bar
c^{\bar m, \bar r}_{n,l}\,\,
 {{\rm L}_{0} [ t_{\bar m, \bar r} / t_{\bar m+1, \bar r}  ]
    \over t_{\bar m+1, \bar r}  }\,. \cr} \equn
$$
where
$\Slash{\!P} = \Slash{\!\tilde P} = \mathbb{I}$ in the $\NeqOne$ triangle
coefficients $c^{m,r}_{n,l}$ and $\bar c^{\bar m, \bar r}_{n,l}$\,.
For the scalar loop contribution, the $\Kz$ functions and $\Lz$ functions are
just equal to one-third of the $\NeqOne$ contributions, as can be
easily demonstrated from our recursion, so that we
write the full amplitudes as
$$\hspace{0.1cm}
\eqalign{ A^{[0]}(1^-,2^-,\cdots,  l^-,l+1^+,\cdots  n^+ )
\;=\;&\,{1 \over 3}\, A^{\,\NeqOne}(1^-,2^-,\cdots, l^-,l+1^+,\cdots  n^+ )
\cr &\hspace{-4.65cm}\;+\;\frac13\sum_{m,r}^{}\,\, c^{m,r}_{n,l}\,\,
 { \Lzz [ t_{m,r} / t_{m+1,r}] \over t_{m+1,r}^3  }
 \;+\; \frac13\sum_{\bar m, \bar r}\,\, \bar c^{\bar m, \bar r}_{n,l}\,\,
 { \Lzz[ t_{\bar m, \bar r} / t_{\bar m+1, \bar r}  ] \over t_{\bar m+1,
\bar r}^3  } \;+\; \hbox{rational}\,, \cr} \equn
$$
where now $\Slash{\!P} = \Slash{k_m}\Slash {\!K_{L}}$\,,
$\Slash{\!\tilde P} = \Slash {\!K_{L}}\Slash {k_m}$ for the triangle
coefficients $c^{m,r}_{n,l}$ and $\bar c^{\bar m, \bar r}_{n,l}$\,.
In these expressions, $c^{m,r}_{ n,l}$ is the coefficient of the triangle function,
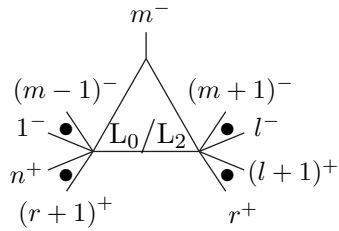
\vspace{-0.5cm}
\begin{center}
\begin{picture}(100,100)(0,0)
\Line(30,30)(70,30) \Line(30,30)(50,65) \Line(50,65)(70,30)
\Line(30,30)(20,45) \Line(30,30)(13,23) \Line(30,30)(13,37) \Line(30,30)(20,15)
\Line(70,30)(80,45) \Line(70,30)(87,37) \Line(70,30)(87,23) \Line(70,30)(80,15)
\Line(50,65)(50,75)
\Text(52,83)[]{\footnotesize$m^-$}

\Text(20,53)[]{\footnotesize$(m-1)^-$}
\Text(86,53)[]{\footnotesize$(m+1)^-$}
\Text(5,22)[]{\footnotesize$n^+$}
\Text(106,22)[]{\footnotesize$(l+1)^+$}
\Text(87,7)[]{\footnotesize$r^+$}
\Text(20,7)[]{\footnotesize${(r+1)}^+$}
\Text(7,40)[]{\footnotesize$1^-$}
\Text(96,40)[]{\footnotesize$l^-$}
\Text(20,21)[]{\textcolor{blue}{$\bullet$}}
\Text(81,38)[]{\textcolor{red}{$\bullet$}}
\Text(81,21)[]{\textcolor{blue}{$\bullet$}}
\Text(20,38)[]{\textcolor{red}{$\bullet$}}
\Text(51,36)[]{\small ${\rm L}_0\big/{\rm L}_2$}
\end{picture}
\end{center}
\vspace{-0.4cm}
while the  $\bar c^{\bar m, \bar r}_{ n,l}$ is the coefficient of the
triangle function,
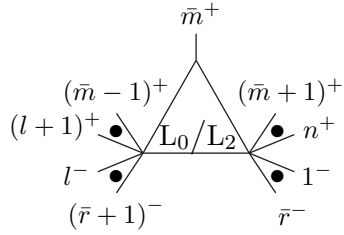
\vspace{-0.5cm}
\begin{center}
\begin{picture}(100,100)(0,0)
\Line(30,30)(70,30) \Line(30,30)(50,65) \Line(50,65)(70,30)
\Line(30,30)(20,45) \Line(30,30)(13,23) \Line(30,30)(13,37) \Line(30,30)(20,15)
\Line(70,30)(80,45) \Line(70,30)(87,37) \Line(70,30)(87,23) \Line(70,30)(80,15)
\Line(50,65)(50,75)
\Text(52,83)[]{\footnotesize$\bar m^+$}

\Text(20,53)[]{\footnotesize$(\bar m-1)^+$}
\Text(86,53)[]{\footnotesize$(\bar m+1)^+$}
\Text(5,22)[]{\footnotesize$l^-$}
\Text(96,22)[]{\footnotesize$1^-$}
\Text(87,7)[]{\footnotesize$\bar r^-$}
\Text(20,7)[]{\footnotesize${(\bar r+1)}^-$}
\Text(-3,40)[]{\footnotesize$(l+1)^+$}
\Text(97,40)[]{\footnotesize$n^+$}

\Text(20,21)[]{\textcolor{blue}{$\bullet$}}
\Text(81,38)[]{\textcolor{red}{$\bullet$}}
\Text(81,21)[]{\textcolor{blue}{$\bullet$}}
\Text(20,38)[]{\textcolor{red}{$\bullet$}}
\Text(51,36)[]{\small ${\rm L}_0\big/{\rm L}_2$}
\end{picture}
\end{center}
\vspace{-0.4cm}
Note that the coefficients $\bar c^{\bar m,\bar r}_{ n,l}$ are
simply related to the coefficients $c^{m,r}_{ n,l}$ by parity and relabelling.
Thus, we need only obtain the  $c^{m,r}_{ n,l}$ coefficients to have
a solution.

The coefficients of the $\Kz$ functions satisfy the same recursion
relations as the tree amplitudes with the starting point being
a five-point tree amplitude. This
gives us the coefficients of the $\Kz$ functions to be proportional
to tree amplitudes,
$$
\eqalign{\;\;\; a_1  = a_2 ={1 \over 2 }
\Atree(1^-,\ldots,l^-,l+1^+,\ldots,n^+)\,,} \equn
$$
The formulae for the tree amplitudes may be found in ref.~\cite{SplitHelicity}.

As a warm up before dealing with the all-$n$
point cases of triangle coefficients, we first consider specific
and detailed examples of how to recursively derive the coefficients of
the split-helicity triangle integral functions.

\subsection{Seven Points}

The first example we consider is the googly NMHV amplitude $A(1^-,2^-,3^-,4^-,5^+,6^+,7^+)$  and the coefficient
$c_7=c_7(7^+,1^-;2^-;3^-,4^-,5^+,6^+)$ which is associated with the integral
functions $\Lz(-t_{7,1}/(-t_{7,2}))$ and $\Lzz(-t_{7,1}/(-t_{7,2}))$\,,
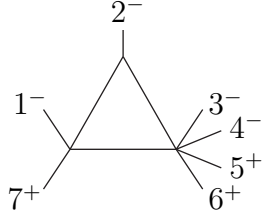
\vspace{-0.6cm}
\begin{center}
\begin{picture}(100,100)(0,0)
\Line(30,30)(70,30)
\Line(30,30)(50,65)
\Line(50,65)(70,30)

\Line(30,30)(20,45)
\Line(30,30)(20,15)

\Line(70,30)(80,45)
\Line(70,30)(87,37)
\Line(70,30)(87,23)
\Line(70,30)(80,15)

\Line(50,65)(50,75)

\Text(52,83)[]{$2^-$}

\Text(15,48)[]{$1^-$}
\Text(13,12)[]{${7}^+$}

\Text(89,48)[]{$3^-$}
\Text(97,39)[]{$4^-$}
\Text(97,24)[]{$5^+$}
\Text(89,12)[]{$6^+$}

\Text(-20,55)[]{$$}
\Text(50,43)[]{$$}
\end{picture}
\end{center}
\vspace{-0.6cm}
In the recursion we shift $p_4=p_4(z)$\,, $p_5=p_5(z)$ according to
$$
\hspace{1cm}
\eqalign{
\tilde\lambda_4&\;\rightarrow\;\tilde\lambda_4\;-\;z\tilde\lambda_5\,,\cr
\lambda_5&\;\rightarrow\;\lambda_5\;+\;z\lambda_4\,.
}\equn\label{Shift45}
$$
The momenta $p_4^2(z)$ and $p_5^2(z)$ remain on-shell, {\it i.e.}
$p_4^2(z)=0$ and $p_5^2(z)=0$ and momentum conservation is unchanged.
The shift we consider here satisfies the two criteria for valid shifts
given in \sec{LoopRecursionSection}.  We will demonstrate this for
all-$n$ in section~\ref{allnsection}.

Under this shift, the contributing factorisations
are given by the collinear limits for $p_3\parallel p_4$\,,
$$
\hspace{0.35cm}
\eqalign{
c_7(7^+,1^-;2^-;3^-,4^-,5^+,6^+)\ \inlimit^{3 \parallel 4}\
{\rm Split}_+(3^-,4^-)\,c_6(7^+,1^-;2^-;(3+4)^-,5^+,6^+)\,,
}\equn
$$
and for $p_5\parallel p_6$\,,
$$
\hspace{0.35cm}
\eqalign{
c_7(7^+,1^-;2^-;3^-,4^-,5^+,6^+)\ \inlimit^{5 \parallel 6}\
{\rm Split}_-(5^+,6^+)\,c_6(7^+,1^-;2^-;3^-,4^-,(5+6)^+)\,.
}\equn
$$
Thus the coefficient $c_7(z)$ has only two simple poles in $z$\,,
corresponding to the two collinear factorisations.
The six-point coefficients that go into the collinear factorisations
$p_3\parallel p_4$ and $p_5\parallel p_6$ can both be derived by
recursion from the five-point coefficient as we have seen in
section~\ref{recursion}.

By recursion we then calculate the coefficient $c_7$ as the residue of
$-c_7(z)/z$ at the  poles $z_1=[3\,4]/[3\,5]$ and
$z_2=-\la5\,6\ra/\la4\,6\ra$\,,
$$
c_7\;=\;c_7(0)\;=\;-\sum_{\alpha=1,2}{\rm Res}_{z=z_\alpha}\frac{c_7(z)}{z}\,,
\equn$$
{\it i.e.} the contribution from $z_1$ is given by
$$
c_7|_{z_1}\;=\;c_6(7^+,1^-;2^-;\hat K_{3,4}^-\,,\hat
5^+,6^+)\,\frac{i}{s_{34}}\,A(\hat 3^-,4^-,(-\hat K_{3,4})^-)\,.
\equn$$ When inserting the form eq.~(\ref{C6p}) for
$c_6(7^+,1^-;2^-;\hat K_{3,4}^-,\hat 5^+,6^+)$ we can write
$c_7|_{z_1}$ as a double shift of the five-point googly coefficient
$c_5$\,,
$$
c_7|_{{z_1}}\;=\;c_5(7^+,1^-;2^-;\hat{\hat{K}}_{3,4}^-\,,\hat{K}_{\hat5,6}^+\,)\frac{i}{s_{\hat
56}}A(\hat{\hat{5}}^+,6^+,(-\hat{K}_{\hat5,6})^-)\frac{i}{s_{34}}\,
A(3^-,\hat 4^-,(-\hat K_{3,4})^+)\,.\equn
$$
The expression for $c_7|_{z_1}$ is thus given by a sequence of shifts
of the five-point googly coefficient $c_5$\,. The first shift adds a
positive helicity leg, while the second shift adds a negative helicity
leg.  As we increase the number of legs the repeated shifts
will be very useful as a way of organising the results.
In the following we will call shifts that add a positive helicity leg
a `plus shift' and a shift that adds a negative helicity leg a
`minus shift'. The double hat in the above equation expresses that
we are shifting a leg twice.

\noindent
Explicitly we find for $c_7|_{z_1}$\,,
$$\hspace{-3.9cm}
  c_7|_{z_1}\;=\;-i\frac{[5|K_{3,5}K_{3,6}P|7]\,[5|K_{3,5}K_{3,6}\tilde P|7]\, [5|K_{3,5}K_{3,6}[k_2,K_L]|7]}{[3|K_{3,5}|6\ra\la6|K_{3,6}|2]\spb7.1\spb1.2\spb3.4\spb4.5\,t_{3,5}\,t_{3,6}}\,.\equn
$$
\noindent
The contribution $c_7|_{z_2}$ is given by a minus shift followed
by a plus shift of $c_5$\,,
$$\hspace{0.5cm}
\eqalign{
  c_7|_{z_2}&\;=\;c_5(7^+,1^-;2^-;\hat{K}_{3,\hat{4}}^-\,,\hat{\hat{K}}_{5,6}^+\,)\frac{i}{s_{3\hat{4}}}\,
  A(3^-,\hat{\hat{4}}^-,(-\hat{K}_{3,\hat{4}})^+)\frac{i}{s_{56}}A(\hat{5}^+,6^+,(-\hat{K}_{5,6})^-)\,,\cr
  &\;=\;-i\frac{\la 4|K_{4,6}P|7]\,\la 4|K_{4,6}\tilde P|7]\,\la 4|K_{4,6}[k_2,K_L]|7]}{\BBRQ 3 {K_{4,6}}6\,\spb7.1\spb1.2 \spb2.3\spa4.5\spa5.6\, t_{4,6}}\,.}
\equn
$$
We find it useful to keep track of the various contributions to the
integral coefficients via representations of these as different paths
in a helicity diagram. Such a diagram provides a pictorial expression
for the factorisation structure of a massive corner in a
coefficient.

In our conventions the point $(1,1)$ will correspond to
a corner with one positive helicity leg and one negative, {\it i.e.}
$(p,n) = (1,1)$\,. The point (1,1) is thus associated with the
googly five-point coefficient $c_5$\,.  The path going one step up and
right ($\nearrow$) corresponds to a plus shift, while a path
going one up and left ($\nwarrow$) should be associated with a
minus shift. Hence the coefficient $c_5$ is related by a plus shift to the
part given by the googly six-point coefficient represented by $(1,2)$\,.
By a subsequent minus shift one can then get the part $c_7|_{z_1}$ of
the seven-point coefficient corresponding to the point $(2,2)$ in the
diagram. Another possibility is to go from the coefficient $c_5$
to the NMHV six-point coefficient represented by $(2,1)$ by a minus shift and
subsequently by a plus shift to get the part $c_7|_{z_2}$\,.

The helicity diagrams corresponding to these two
different shift paths are depicted below.
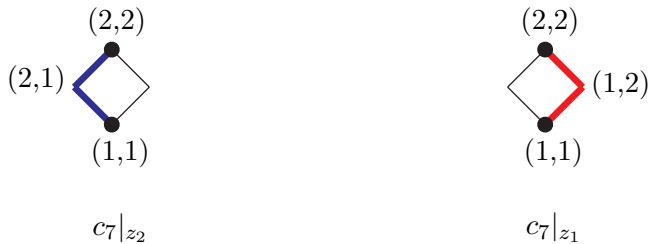
\vspace{-0.9cm}\begin{center}
\begin{picture}(60,80)(80,-30)\rotatebox{-45}{
\Line(40,40)(60,40)
\Line(40,40)(40,60)
\Line(60,40)(60,60)
\Line(40,60)(60,60)
\SetColor{Blue}
\SetWidth{2.5}
\Line(40,40)(40,60)
\Line(40,40)(60,40)
\Text(70,35)[]{\rotatebox{45}{\small (1,1)}}
\Text(28,32)[]{\rotatebox{45}{\small (2,1)}}
\Text(34,69)[]{\rotatebox{45}{\small (2,2)}}
\Text(90,15)[]{\rotatebox{45}{\small $c_7|_{z_2}$}}
\SetColor{Black}
\Vertex(60,40){3}
\Vertex(40,60){3}
}
\end{picture}
\begin{picture}(60,80)(-20,-30)
\rotatebox{-45}{
\Line(40,40)(60,40)
\Line(40,40)(40,60)
\Line(60,40)(60,60)
\Line(40,60)(60,60)
\SetColor{Red}
\SetWidth{2.5}
\Line(40,60)(60,60)
\Line(60,40)(60,60)
\Text(70,35)[]{\rotatebox{45}{\small (1,1)}}
\Text(70,71)[]{\rotatebox{45}{\small (1,2)}}
\Text(34,69)[]{\rotatebox{45}{\small (2,2)}}
\SetColor{Black}
\Text(90,15)[]{\rotatebox{45}{\small $c_7|_{z_1}$}}
\Vertex(60,40){3}
\Vertex(40,60){3}
}
\end{picture}
\end{center}
\vspace{0.8cm}
A diagrammatic approach was also used at tree level to
represent different contributions to split-helicity amplitudes
in~\cite{SplitHelicity}.

We conclude this subsection by noting that the complete expression for
the googly NMHV coefficient $c_7$ is given by the sum over shifts of
the coefficient $c_5$\,. Expressed in terms of the helicity diagram, the
coefficient is given as a sum over paths.
We have checked that the expression for the seven-point-googly NMHV
coefficient in the $\NeqOne$ case
$$
c_7 \;=\; c_7|_{z_1}\;+\; c_7|_{z_2}\,,
\equn
$$
agrees with the results given previously in ref.~\cite{BBDPSQCD}.


\subsection{Eight Points}

It is also worthwhile to work through an eight-point googly NMHV triangle
coefficient since this case has some non-trivial aspects in its
recursive build up which will be important for us when considering the
generic $n$-point triangle coefficients.

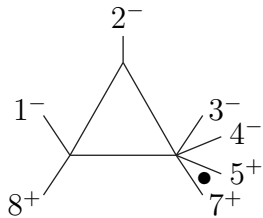
\vspace{-0.6cm}
\begin{center}
\begin{picture}(100,100)(0,0)
\Line(30,30)(70,30)
\Line(30,30)(50,65)
\Line(50,65)(70,30)

\Line(30,30)(20,45)
\Line(30,30)(20,15)

\Line(70,30)(80,45)
\Line(70,30)(87,37)
\Line(70,30)(87,23)
\Line(70,30)(80,15)

\Line(50,65)(50,75)

\Text(52,83)[]{$2^-$}

\Text(15,48)[]{$1^-$}
\Text(13,12)[]{${8}^+$}

\Text(89,48)[]{$3^-$}
\Text(97,39)[]{$4^-$}
\Text(97,24)[]{$5^+$}
\Text(81,21)[]{\textcolor{blue}{$\bullet$}}
\Text(89,12)[]{$7^+$}

\Text(-20,55)[]{$$}
\Text(50,43)[]{$$}
\end{picture}
\end{center}
\vspace{-0.6cm}
In this case we use the shift
$$
\eqalign{
\tilde\lambda_4&\;\rightarrow\;\tilde\lambda_4\;-\;z\tilde\lambda_5\,,\cr
\lambda_5&\;\rightarrow\;\lambda_5\;+\;z\lambda_4\,.}
\equn
$$
Again, this shift satisfies our criteria for valid shifts,
as we will show below in section~\ref{allnsection}.

The eight-point coefficient
$c_8=c(8^+,1^-;2^-;3^-,4^-,5^+,6^+,7^+)$ is given by the sum
$$\eqalign{
c_8&\;=\;c_8|_{z_1}\;+\;c_8|_{z_2}\,,\cr
c_8|_{z_1}&\;=\;c(8^+,1^-;2^-;\hat K_{3,4}^-\,,\hat 5^+,6^+,7^+)\,
\frac{i}{s_{34}}\,A(3^-,\hat 4^-,(-\hat K_{3,4})^+)\,,\cr
c_8|_{z_2}&\;=\;c(8^+,1^-;2^-;3^-,4^-,\hat{K}_{5,6}^+\,,7^+)
\frac{i}{s_{56}}A(\hat{5}^+,6^+,(-\hat{K}_{5,6})^-)\,.}
\equn
$$

When inserting the expression for the seven-point coefficients into the
recursion for $c_8$\,, one finds that it is related to the coefficient
$c_5$ by three different sequences of shifts: 1) two plus shifts and a
subsequent minus shift, 2) a plus shift, a minus shift and a plus shift
and 3) by a minus shift and subsequently two plus shifts. Schematically
this means that the eight-point coefficient is given by
$$
\hspace{0.7cm}\eqalign{
  c_8\;=\;c_8^1\;+\;c_8^2\;+\;c_8^3
  &\;=\;c_5(+---+)\times A(++-)\times A(++-)\times A(--+)\cr
      &\;\hspace{0.06cm}+\;c_5(+---+)\times A(++-)\times A(--+)\times A(++-)\cr
      &\;\hspace{0.06cm}+\;c_5(+---+)\times A(--+)\times A(++-)\times A(++-)\,.
}\equn$$

\begin{center}
\begin{picture}(60,80)(160,-30)
\rotatebox{-45}{\Line(40,40)(60,40)
\Line(40,40)(40,60)
\Line(60,40)(60,60)
\Line(40,60)(60,60)
\Line(40,60)(60,60)
\Line(40,60)(40,80)
\Line(60,60)(60,80)
\Line(40,80)(60,80)
\SetColor{Red}
\SetWidth{2.5}
\Line(40,80)(60,80)
\Line(60,40)(60,80)
\Text(70,35)[]{\rotatebox{45}{\small (1,1)}}
\Text(90,15)[]{\rotatebox{45}{\small $c_8^1$}}
\Text(76,64)[]{\rotatebox{45}{\small (1,2)}}
\Text(73,86)[]{\rotatebox{45}{\small (1,3)}}
\Text(30,85)[]{\rotatebox{45}{\small (2,3)}}
\SetColor{Black}
\Vertex(60,40){3}
\Vertex(40,80){3}
}
\end{picture}
\begin{picture}(60,80)(60,-30)
\rotatebox{-45}{
\Line(40,40)(60,40)
\Line(40,40)(40,60)
\Line(60,40)(60,60)
\Line(40,60)(60,60)
\Line(40,60)(60,60)
\Line(40,60)(40,80)
\Line(60,60)(60,80)
\Line(40,80)(60,80)
\SetColor{Red}
\SetWidth{2.5}
\Line(40,60)(60,60)
\Line(60,40)(60,60)
\Line(40,60)(40,80)
\Text(70,35)[]{\rotatebox{45}{\small (1,1)}}
\Text(90,15)[]{\rotatebox{45}{\small $c_8^2$}}
\Text(74,62)[]{\rotatebox{45}{\small (1,2)}}
\Text(25,56)[]{\rotatebox{45}{\small (2,2)}}
\Text(30,85)[]{\rotatebox{45}{\small (2,3)}}
\SetColor{Black}
\Vertex(60,40){3}
\Vertex(40,80){3}
}
\end{picture}
\begin{picture}(60,80)(-50,-30)
\rotatebox{-45}{\Line(40,40)(60,40)
\Line(40,40)(40,60)
\Line(60,40)(60,60)
\Line(40,60)(60,60)
\Line(40,60)(60,60)
\Line(40,60)(40,80)
\Line(60,60)(60,80)
\Line(40,80)(60,80)
\SetColor{Blue}
\SetWidth{2.5}
\Line(40,40)(40,80)
\Line(40,40)(60,40)
\Text(70,35)[]{\rotatebox{45}{\small (1,1)}}
\Text(90,15)[]{\rotatebox{45}{\small $c_8^3$}}
\Text(27,33)[]{\rotatebox{45}{\small (2,1)}}
\Text(27,57)[]{\rotatebox{45}{\small (2,2)}}
\Text(30,85)[]{\rotatebox{45}{\small (2,3)}}
\SetColor{Black}
\Vertex(60,40){3}
\Vertex(40,80){3}
}
\end{picture}
\end{center}
\vspace{0.6cm} Again we can picture the various contributions to the
eight-point coefficient in helicity dia\-grams.

\subsection{All-$n$ Triangle Coefficient}\label{allnsection}

We now discuss the recursive calculation of the
$n$-point triangle coefficients $c_{n,l}^{m,r}$ for the cases of a
chiral $\NeqOne$ multiplet or scalar in the loop.  As before
we calculate both of these cases simultaneously, using the unified notation
of having $P$ and $\tilde P$\,, defined in \eqn{PDef}, to keep track of the
two different types of loop contributions.

A generic $n$-point triangle function in the split-helicity
amplitude is diagrammatically,
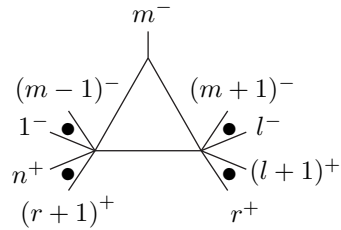
\vspace{-0.5cm}
\begin{center}
\begin{picture}(100,100)(0,0)
\Line(30,30)(70,30) \Line(30,30)(50,65) \Line(50,65)(70,30)
\Line(30,30)(20,45) \Line(30,30)(13,23) \Line(30,30)(13,37) \Line(30,30)(20,15)
\Line(70,30)(80,45) \Line(70,30)(87,37) \Line(70,30)(87,23) \Line(70,30)(80,15)
\Line(50,65)(50,75)
\Text(52,83)[]{\footnotesize$m^-$}

\Text(20,53)[]{\footnotesize$(m-1)^-$}
\Text(86,53)[]{\footnotesize$(m+1)^-$}
\Text(5,22)[]{\footnotesize$n^+$}
\Text(106,22)[]{\footnotesize$(l+1)^+$}
\Text(87,7)[]{\footnotesize$r^+$}
\Text(20,7)[]{\footnotesize${(r+1)}^+$}
\Text(7,40)[]{\footnotesize$1^-$}
\Text(96,40)[]{\footnotesize$l^-$}
\Text(20,21)[]{\textcolor{blue}{$\bullet$}}
\Text(81,38)[]{\textcolor{red}{$\bullet$}}
\Text(81,21)[]{\textcolor{blue}{$\bullet$}}
\Text(20,38)[]{\textcolor{red}{$\bullet$}}
\end{picture}
\end{center}
\vspace{-0.4cm}
At least one of the three-legs must be massless because otherwise the
triple cut would vanish. (The case where the massless leg is of positive
helicity can be obtained by conjugation.)

Consider first a recursion on the right-most corner of the triangle.
We shift legs $l$ and $l+1$\,,
$$\eqalign{
\tilde\lambda_l&\;\rightarrow\;\tilde\lambda_l\;-\;z\tilde\lambda_{l+1}\,,\cr
\lambda_{l+1}&\;\rightarrow\;\lambda_{l+1}\;+\;z\lambda_l\,.} \equn
$$
This corresponds the shifts considered in our previous examples.
The momenta $p_l^2(z)$ and $p_{l+1}^2(z)$ remain
on-shell, {\it i.e.} $p_l^2(z)=0$ and $p_{l+1}^2(z)=0$ and the
total momentum remains zero.

To check if this shift is valid according to the criteria
given in section~\ref{rerelfintco}, we need to
consider the behaviour of the tree amplitudes isolated by the cut,
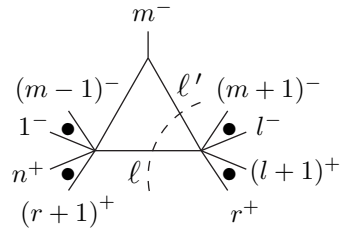
\vspace{-0.8cm}
\begin{center}
\begin{picture}(100,100)(0,0)
\Line(30,30)(70,30) \Line(30,30)(50,65) \Line(50,65)(70,30)
\Line(30,30)(20,45) \Line(30,30)(13,23) \Line(30,30)(13,37) \Line(30,30)(20,15)
\Line(70,30)(80,45) \Line(70,30)(87,37) \Line(70,30)(87,23) \Line(70,30)(80,15)
\Line(50,65)(50,75)
\Text(52,83)[]{\footnotesize$m^-$}

\Text(20,53)[]{\footnotesize$(m-1)^-$}
\Text(96,53)[]{\footnotesize$(m+1)^-$}
\Text(5,22)[]{\footnotesize$n^+$}
\Text(106,22)[]{\footnotesize$(l+1)^+$}
\Text(87,7)[]{\footnotesize$r^+$}
\Text(20,7)[]{\footnotesize${(r+1)}^+$}
\Text(7,40)[]{\footnotesize$1^-$}
\Text(96,40)[]{\footnotesize$l^-$}
\Text(20,21)[]{\textcolor{blue}{$\bullet$}}
\Text(81,38)[]{\textcolor{red}{$\bullet$}}
\Text(81,21)[]{\textcolor{blue}{$\bullet$}}
\Text(20,38)[]{\textcolor{red}{$\bullet$}}
\Text(45,22)[]{${\ell}$}
\Text(66,55)[]{$\ell\,'$}
\DashCArc(80,20)(30,110,190){4}
\end{picture}

\end{center}
\vspace{-0.4cm}
The tree amplitude on the left-hand side of the cut is,
$$
A^{\rm tree}((-\ell_{f/s})^{\pm},(r+1)^+,\ldots ,n^+,1^-,\ldots,
       m^-,(-\ell\,'^{\mp}_{f/s}))\,,
\equn\label{LeftCut}
$$
and on the right-hand side is,
$$
A^{\rm tree}(\ell\,'^{\pm}_{f/s},(m+1)^-,\ldots ,l^-,(l+1)^+,
      \ldots,r^+,\ell_{f/s}^{\mp})\,,
\equn\label{RightCut}
$$
The subscripts $f/s$ signifies either a scalar or
a fermion in the loop.

Note that the shift does not modify the loop momenta appearing
in the cut; if we take
$\ell$ as the independent variable, $\ell\,'$ will not depend on $z$\,, as
it is a function of the sum of the momenta in the tree,
$$
-\ell\,'\;=\;p_{m+1}\;+\ldots+\;p_{l}(z)\;+\;p_{l+1}(z)\;+\ldots+\;\ell\,,
\equn
$$
which is independent of $z$ because of \eqn{BasicShift}.
Thus the $z$-dependence cannot enter into the tree amplitude (\ref{LeftCut})
on the left-hand side of the cut.

Now consider the tree-amplitude (\ref{RightCut}) on the
right-hand-side of the cut, which is the one containing the shifted
legs.  The shifted amplitude can be written as a sum of terms where
each contains only a single pole in $z$ corresponding to a
factorisation of the tree amplitude.  Only factorisations where
loop-momentum flows between the two factor tree amplitudes can
introduce a problematic pole,
$$
\frac{i}{(\ell+\ldots+p_l(z))^2}\,.
\equn\label{TreePole}
$$
However there is no such factorisation of the tree,
as the only potential factorisation that could contribute,
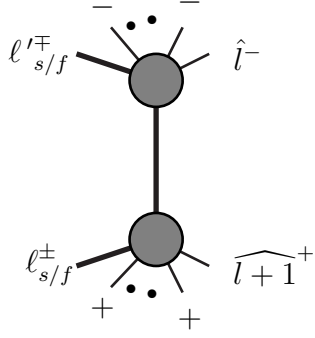
\vspace{-0.3cm}
\begin{center}
\begin{picture}(100,120)(0,0)
\SetWidth{2}
\Line(50,20)(50,80)
\Line(50,20)(20,10)
\Line(50,80)(20,90)
\SetWidth{1}
\Line(50,20)(70,10)
\Line(50,80)(70,90)
\Line(50,20)(60,0)
\Line(50,80)(60,100)
\Line(50,20)(33,2)

\Line(50,80)(33,100)

\GCirc(50,20){10}{0.5}
\GCirc(50,80){10}{0.5}

\Text(10,10)[]{$\ell_{s/f}^{\pm}$}
\Text(6,90)[]{${\ell\,'}_{s/f}^{\mp}$}
\Text(85,90)[]{$\hat{l}^-$}
\Text(95,10)[]{$\widehat{l+1}^+$}
\Text(63,-10)[]{$+$}
\Text(63,110)[]{$-$}

\Text(30,-6)[]{$+$}
\Text(30,108)[]{$-$}

\Text(41,-1)[]{$^\bullet$}
\Text(41,101)[]{$_\bullet$}
\Text(49,-3)[]{$^\bullet$}
\Text(49,103)[]{$_\bullet$}
\end{picture}
\end{center}
\vspace{0.4cm}
vanishes because the helicity structure forces at least one of the two
tree amplitudes appearing in the factorization to vanish, assuming
there are at least a total of three external legs (not counting the
ones carrying loop momentum).  The vanishing of this pole contribution
is related to the lack of box integrals in these amplitudes. This
shows that the second criterion of
\sec{IntegralCoeffRecursionSubsection} that all loop momentum
dependent poles are unmodified by the shift is satisfied.

The first criterion for the shift that the coefficients $c(z)$
vanish when $z$ goes to infinity is also satisfied.
This follows from the vanishing of the
tree-amplitude in the cut as $z$ goes to infinity and
because all $z$-dependence can be pulled out of the loop momentum
integral. The limit $|z|\rightarrow\infty$ then commutes trivially with the
integration and we conclude that the coefficient times the
integral vanishes in this limit.  This proves that our recursion is valid
and that there are no boundary terms.

It is worth noting that since the shift comes out of
the integration, all logarithms will be unaffected
under the shifts, so that $\ln(t_{i,j})\rightarrow
\ln(t_{i,j})$.

Since we have proven that our shift is valid for all integral
coefficients in the split-helicity amplitudes, we may now consider the
general recursion. With our choice of shift,
the only relevant factorisations are given by the
collinear limits for $p_{l-1}\parallel p_l$\,,
$$
\hspace{0.3cm}
\eqalign{
&c^{m,r}_{n,l}\ \inlimit^{l-1 \parallel l}\ {\rm Split}_+
\big((l-1)^-,l^-\big)\times c^{m,r}_{n-1,l-1}\,, \cr
}
\equn
$$
and for $p_{l+1}\parallel p_{l+2}$\,,
$$
\hspace{0.9cm}
\eqalign{
&c^{m,r}_{n,l} \ \inlimit^{l+1 \parallel l+2}\ {\rm
Split}_-\big((l+1)^+,(l+2)^+\big)\times c^{m,r}_{n-1,l}\,. \cr
}
\equn
$$

The coefficient $c_{n,l}^{m,r}$ can now, as in the examples,
be computed as the residues of $-c_{n,l}^{m,r}(z)/z$ at its two
collinear poles
$z_1=[l-1,l]/[l-1,l+1]$ and $z_2=-\la l+1,l+2\ra/\la l,l+2\ra$\,,
$$
c_{n,l}^{m,r}\;=\;c_{n,l}^{m,r}(0)\;=\;-\sum_{i=1,2}{\rm Res}_{z=z_i}
\frac{c_{n,l}^{m,r}(z)}{z}\,,
 \equn
$$
which gives
$$
\hspace{0.8cm}\eqalign{
c_{n,l}^{m,r}&\;=\;c_{n,l}^{m,r}|_{z_1}\;+\;c_{n,l}^{m,r}|_{z_2}\,,\cr
c_{n,l}^{m,r}|_{z_1}&\;=\;c_{n-1,l-1}^{m,r}
\big(...,\hat{K}_{l-1,l}^-\,,\widehat{(l+1)}^+,...\big)\,
\frac{i}{s_{l-1,l}}\,A\big((l-1)^-,\hat
l^-,(-\hat K_{l-1,l})^+\,\big)\,,\cr
c_{n,l}^{m,r}|_{z_2}&\;=\;c_{n-1,l}^{m,r}
\big(...,\hat{l}^-,\hat{K}_{l+1,l+2}^+\,,...\big)\,
\frac{i}{s_{l+1,l+2}}\,A\big(\widehat{(l+1)}^+,(l+2)^+,(-\hat
K_{l+1,l+2})^-\big)\,. }
\equn
$$
Thus, we have succeeded in expressing $n$-point coefficients
in terms of $(n-1)$-point
coefficients. The analogous steps can be applied to the left massive
corner. By repeated application of the above steps $c_{n,l}^{m,r}$ can
be written in terms of shifts of the basic coefficient $c_5$\,. On
each recursion the coefficient is a sum of two terms: one based on the
$(n-1)$-point coefficient with one less negative helicity and the
other based upon the $n-1$ point coefficient with one less positive
helicity.  One can think of the recursion as leading via a ``path''
back to the coefficient of the five-point amplitude where the massive
corner contains exactly one negative and one positive helicity leg.
The coefficients $c_{n,l}^{m,r}$ are then given by the sum,
$$
\eqalign{
\label{sumoverpath}
c_{n,l}^{m,r}\;=\;\sum_{P_L,P_R}\,\,T_{P_L,P_R}\,. }
\equn\label{EQsumPaths}
$$
The $T_{P_L,P_R}$ term are generated
from the same starting point; the five-point googly
triangle coefficient $c(5^+,1^-;2^-;3^-,4^+)$ by
sequences of plus/minus shifts on the left and right
massive corners independently.
The sequence of shifts
is encoded in the
arguments $P_L$ and $P_R$ and correspond to a solution of a set of
parameters $\alpha_i,\,\rho_i,\,\beta_{i'},\,\sigma_{i'}$ in
a helicity diagram.
Two examples of paths for a massive corner are, 
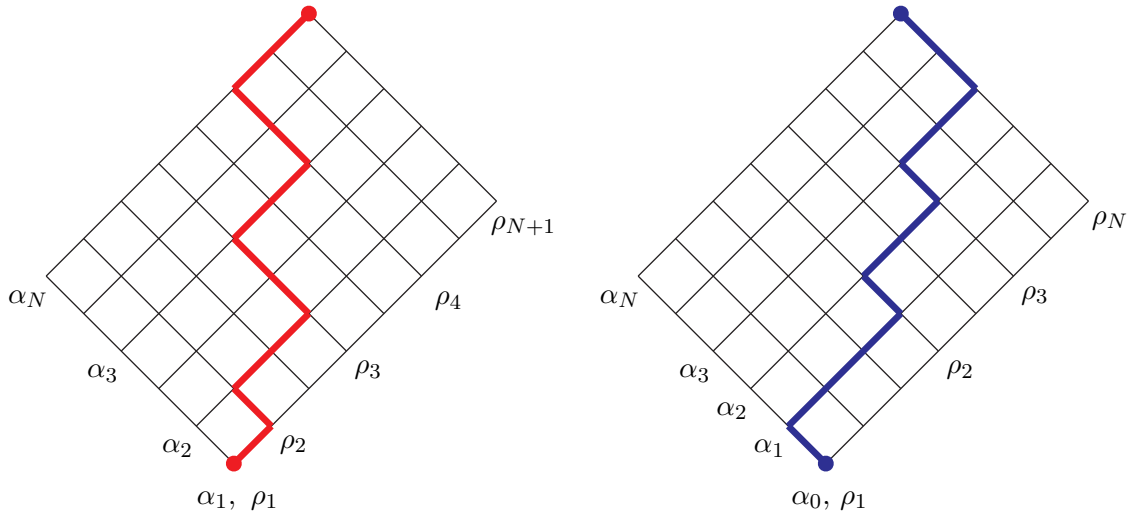
\vspace{-1.4cm}
\begin{center}
\begin{picture}(100,160)(120,0)
\rotatebox{-45}{
\Line(20,40)(120,40) \Line(20,60)(120,60) \Line(20,80)(120,80)
\Line(20,100)(120,100) \Line(20,120)(120,120)
\Line(20,140)(120,140) \Line(20,160)(120,160)
\Line(20,180)(120,180)
\Line(120,40)(120,180) \Line(100,40)(100,180) \Line(80,40)(80,180)
\Line(60,40)(60,180) \Line(40,40)(40,180) \Line(20,40)(20,180)
\SetColor{Red} \SetWidth{2.5}
\Line(120,40)(120,60) \Line(120,60)(100,60) \Line(100,60)(100,100)
\Line(100,100)(60,100) \Line(60,100)(60,140) \Line(60,140)(20,140)
\Line(20,140)(20,180)
\Text(130,30)[]{\small \rotatebox{45}{ $\alpha_1,\,\,\rho_1$}}
\Vertex(120,40){3} \Vertex(20,180){3} \Text(100,28)[]{\small
\rotatebox{45}{ $\alpha_2$}} \Text(60,28)[]{\small \rotatebox{45}{
$\alpha_3$}} \Text(20,28)[]{\small \rotatebox{45}{ $\alpha_N$}}
\Text(40,28)[]{\small \rotatebox{45}{
$$}}
\Text(20,28)[]{\small \rotatebox{45}{
$$}}
\Text(0,28)[]{\small \rotatebox{45}{
$$}}
\Text(-20,28)[]{\small \rotatebox{45}{
$$}}
\Text(-40,28)[]{\small \rotatebox{45}{
$$}}
\Text(130,60)[]{\small \rotatebox{45}{ $\rho_2$}}
\Text(130,100)[]{\small \rotatebox{45}{ $\rho_3$}}
\Text(130,100)[]{\small \rotatebox{45}{
$$}}
\Text(130,120)[]{\small \rotatebox{45}{
$$}}
\Text(133,140)[]{\small \rotatebox{45}{ $\rho_4$}}
\Text(133,160)[]{\small \rotatebox{45}{
$$}}
\Text(133,180)[]{\small \rotatebox{45}{ $\rho_{N+1}$}}
\Text(133,200)[]{\small \rotatebox{45}{
$$}}
}\end{picture}
\begin{picture}(100,160)(0,0)
\rotatebox{-45}{ \Line(20,40)(120,40) \Line(20,60)(120,60)
\Line(20,80)(120,80) \Line(20,100)(120,100) \Line(20,120)(120,120)
\Line(20,140)(120,140) \Line(20,160)(120,160)
\Line(20,180)(120,180)
\Line(120,40)(120,180) \Line(100,40)(100,180) \Line(80,40)(80,180)
\Line(60,40)(60,180) \Line(40,40)(40,180) \Line(20,40)(20,180)
\SetColor{Blue} \SetWidth{2.5}
\Line(120,40)(100,40)
\Line(100,40)(100,100) \Line(100,100)(80,100)
\Line(80,100)(80,140) \Line(80,140)(60,140) \Line(60,140)(60,180)
\Line(60,180)(20,180)
\Text(130,30)[]{\small \rotatebox{45}{ $\alpha_0,\,\rho_1$}}
\Vertex(120,40){3} \Vertex(20,180){3} \Text(100,28)[]{\small
\rotatebox{45}{ $\alpha_1$}} \Text(80,28)[]{\small \rotatebox{45}{
$\alpha_2$}} \Text(60,28)[]{\small \rotatebox{45}{ $\alpha_3$}}
\Text(40,28)[]{\small \rotatebox{45}{
$$}}
\Text(20,28)[]{\small \rotatebox{45}{ $\alpha_N$}}
\Text(0,28)[]{\small \rotatebox{45}{
$$}}
\Text(-20,28)[]{\small \rotatebox{45}{
$$}}
\Text(-40,28)[]{\small \rotatebox{45}{
$$}}
\Text(130,100)[]{\small \rotatebox{45}{ $\rho_2$}}
\Text(130,120)[]{\small \rotatebox{45}{
$$}}
\Text(130,100)[]{\small \rotatebox{45}{
$$}}
\Text(130,120)[]{\small \rotatebox{45}{
$$}}
\Text(130,140)[]{\small \rotatebox{45}{ $\rho_3$}}
\Text(133,160)[]{\small \rotatebox{45}{
$$}}
\Text(130,180)[]{\small \rotatebox{45}{ $\rho_{N}$}}
\Text(133,200)[]{\small \rotatebox{45}{
$$}}
}
\end{picture}
\vspace{2.5cm}
\end{center}
Both paths contribute to the sum in eq.~(\ref{EQsumPaths}) of
a coefficient with $x$ negative helicity legs and $y$ positive
helicity legs on one corner. The coefficient is represented by
the point $(x,y)$\,, with $(x,y)=(l-m,r-l)$\,.
Since each massive leg must be reduced recursively, one sums over two
sets of paths; the paths in a helicity diagram for the left corner and
the paths in a helicity diagram for the right corner.  Following
this logic we have obtained a general solution for the split helicity
with either an $\NeqOne$ multiplet or scalar in the loop.
A path in a helicity diagram can be expressed as,
$$
\eqalign{
&
P_L\; =\; P{[\sigma_j,\beta_{j'}]\,, \;\;\;\quad\quad\hspace{-0.09cm}
j=1,\ldots, N'+\kappa'\,,  \;\;\;\quad\hspace{-0.07cm}  j'=\kappa',\ldots {N'}\,,
}\cr
&P_R\;=\;P[\alpha_i,\rho_{i'}]\,, \;\;\;\quad\quad
i=\kappa,\ldots, N\,,   \;\;\;\quad\hspace{0.88cm}
i'= 1,\ldots,N+\kappa\,,
\cr}
\equn
$$
where $\kappa$, $\kappa'$ and $N$, $N'$ are determined
by the path in the helicity diagram. The parameters $\kappa$,
$\kappa'$ will either take the value $0$ or $1$ depending on
if the path in the helicity diagram starts to the left or to
the right. The parameters $N$ and $N'$ are given roughly by the number 
of corners of the path and can take values between $1$ and the number 
of negative and positive legs whichever is least. In this range only 
those values for $N$ and $N'$ will contribute for which the inequalities 
eq.~(\ref{ineqn}) have solutions.

The boundary parameters for specific $\kappa$, $\kappa'$ and $N$, $N'$ are
$$\hskip 1.5 truecm
\eqalign{
\alpha_{\kappa}&\;=\;m+1\,,\hspace{0.88cm}\alpha_N\;=\;l\,,\hspace{1.16cm}
\rho_{N+\kappa}\;=\;l+1\,,\hspace{1.12cm} \rho_1\;=\;r\,,\cr
\beta_{\kappa'}&\;=\;m-1\,,\hspace{0.8cm}\beta_{N'}\;=\;1\,,\hspace{0.9cm}
\sigma_{N'+\kappa'}\;=\;n\,,\hspace{1.7cm}\sigma_1\;=\;r+1\,.}
\equn
\label{BoundaryConstraints}
$$
For a fixed set of boundary  parameters the remaining
parameters will have to satisfy the following set of inequalities,
$$
\eqalign{
\kappa\;=\;0,1,&\quad\hspace{1cm}
\alpha_i\;<\;\alpha_{i+1},\quad\hspace{1cm} i\;=\;\kappa,...,N-2+\kappa\,,\cr &
\quad\hspace{1.05cm} \rho_i\;>\;\rho_{i+1},\quad\hspace{1.05cm}
i\;=\;1,...,N-1\,,\cr
 \kappa'\;=\;0,1,&\quad\hspace{1.01cm} \beta_j\;>\;\beta_{j+1},\quad\hspace{0.93cm} j\;=\;\kappa',...,N'-2+\kappa'\,,\cr
& \quad\hspace{1.01cm} \sigma_j\;<\;\sigma_{j+1},\quad\hspace{0.93cm}
j\;=\;1,...,N'-1\,.
}
\equn\label{ineqn}
$$
For a solution to the inequalities the
$T_{P_L,P_R}$ takes the following compact form
$$\hspace{0.5cm}
\eqalign{ T_{P_L,P_R}&\;=\; i\,
(-1)^{l+N+N'} \, \frac{\big\la l|
Q_R\,P\,Q_L|1\big\ra\,\big\la l| Q_R\,\tilde
P\,Q_L|1\big\ra\,\big\la l| Q_R[k_m,K_L']Q_L|1\big\ra\,\la
r,r+1\ra} {[1\,2][2\,3]...[l-1,l]\la l,l+1\ra...\la
n-1,n\ra\la n,1\ra}\cr &\hspace{2.1cm}\times\frac{\prod_{i=1}^N
[\alpha_i-1,\alpha_i]\,\prod_{i=2}^N \la\rho_i,\rho_i+1 \ra\,}{\la
\rho_N|K_N|\alpha_{N}-1]\, \prod_{i=1}^{N-1}\la
\rho_i|K_i|\alpha_{i}-1]\,[\alpha_i|\bar K_i|\rho_{i+1}+1\ra} \cr
&\hspace{2.1cm}\times \frac{\prod_{j=1}^{N'} [\beta_j,\beta_j+1]\,
\prod_{j=2}^{N'} \la \sigma_j-1,\sigma_j\ra\,}{\la \sigma_{N'}|
K'_{N'}|\beta_{N'}+1]\, \prod_{j=1}^{N'-1}\la \sigma_j|
K'_j|\beta_{j}+1][\beta_j|\bar K'_j|\sigma_{i+1}-1\ra}\cr
&\hspace{2.1cm}\times
\frac{1}{K^2_N\,K^2_{N'}\,\prod_{i=1}^{N-1}\bar{K}^2_i\,K^2_i\,
\,\prod_{j=1}^{N'-1} {K'}_j^2 \bar{K'}_j^2}\,, }
\equn\label{generalT}$$
where
$$
\hspace{2cm}
\eqalign{
 K_i&\;=\;K_{\alpha_i,\rho_i}\,,\hspace{4cm} K_j'\;=
\;K_{\beta_j,\sigma_j} \,,\cr
 \bar K_i&\;=\;K_{\alpha_i,\rho_{i+1}}\,,
\hspace{3.7cm} \bar K_j'\;=\;K_{\beta_j,\sigma_{j+1}}\,,\cr
 K_R&\;=\;K_{\alpha_{\kappa},\rho_1}\,,\hspace{3.90cm} K_L'\;
=\;K_{\beta_{\kappa'},\sigma_1}\,,\cr
 Q_R&\;=\;K_N \bar K_{N-1}K_{N-1}\,\ldots\,\bar K_1K_1\,,\hspace{0.5cm}
 Q_L\;=\;K_1' \bar K_{1}'\,\ldots\,K_{N'-1}'\bar K_{N'-1}'K_{N'}'\,,\cr
 P&\;=\;k_m K_R\,,1\,,
\hspace{3.65cm}
\tilde P\;=\;K_Rk_m\,,1\,.
}
\equn
$$
In order to calculate a triangle coefficient one uses the above
formula and sum overs the contributions to $T_{P_L,P_R}$ for each solution
of the parameters $\sigma_j,\beta_{j'}$ and $\alpha_i,\rho_{i'}$.
This is equivalent to summing over all possible paths in a helicity diagram.


\section{Results for NMHV Split-Helicity Amplitudes}

In this section we specialise the results of the previous section to the
case where there are exactly three nearest neighbouring negative
helicities in the colour ordering.  For this ``next-to-MHV'' (NMHV)
helicity configuration almost all of the supersymmetric decomposition
of the QCD amplitude is known: the $\NeqFour$ contributions are known
from ref.~\cite{BDKn} as are the contributions from the $\NeqOne$
chiral multiplet~\cite{BBDPSQCD}.  The cut containing scalar loop
contributions are contained in the expressions of the previous section
and in the general formula for $T_{P_L,P_R}$.
This leaves only the rational parts as
undetermined. With an evaluation of these rational functions
parts using, for example, the method
of ref.~\cite{BDKSix} we would have the complete
QCD amplitudes for this helicity configuration.

The expressions for the $\NeqOne$ and scalar loop contributions are
rather similar, whereas that of the $\NeqFour$ multiplet is very
different: the $\NeqFour$ amplitude consists entirely of box
functions.  In contrast, the $\NeqOne$ contribution consists
entirely of the $\Lz$ and $\Kz$ functions while the scalar loop
contribution also contains $\Lzz$ functions.

\subsection{Explicit Results}

Using the generic formula for $T_{P_L,P_R}$ we have obtained the following
compact results for the NMHV helicity configurations:
For the amplitudes with an $\NeqOne$ chiral multiplet running in the loop the
result is,
$$
\eqalign{
A_{n}^{\,\NeqOne\ {\rm chiral}}(1^-,2^-,3^-,4^+,5^+,\cdots, n^+)
\; = \; &
{\Atree \over 2} \,\left( \Kz( s_{n1} ) +\Kz( s_{34} )
\right)
-{i \over 2}
\sum_{r=4}^{n-1} \,\hat d_{n,r}\,
{    \Lz [ t_{3,r} / t_{2,r} ] \over t_{2,r}  }
\cr
&\hspace{-0.6cm}
-{i \over 2}
\sum_{r=4}^{n-2}\, \hat g_{n,r}\,
{    \Lz [ t_{2,r} / t_{2,r+1} ] \over t_{2,r+1}  }
-{i \over 2}
\sum_{r=4}^{n-2}\, \hat h_{n,r}\,
{    \Lz [ t_{3,r} / t_{3,r+1} ] \over t_{3,r+1}  }\,,
\cr}
\equn
$$
\noindent where,
$$\hspace{1cm}
\eqalign{
\hat d_{n,r}\; = \; &
{  \la 3 | {K}_{3,r}{K}_{2,r}  | 1 \ra ^2\,
\la 3 | {K}_{3,r}\big[k_2,{K}_{2,r}\big]{K}_{2,r}  | 1 \ra
\over
 [ 2 | {K}_{2,r}  | r \ra
 [ 2 | {K}_{2,r}  | {r+1} \ra\, \spa3.4\ldots \spa{r-1} .{r}
\spa{r+1}.{r+2}\ldots   \spa{n}.1 \,{t}_{2,r}\,t_{3,r}}\,,
\cr
\hat g_{n,r}\ =\ &
\sum_{j=1}^{r-3}
{\la 3|K_{3,j+3} {K}_{2,j+3}|1\ra^2\,
\la 3|K_{3,j+3}{K}_{2,j+3}\big[k_{r+1},{K}_{2,r}\big]|1\ra
\spa{j+3}.{j+4}
\over
[2|K_{2,j+3}|j+3\ra
[2|K_{2,j+3}|j+4\ra\,
\spa{3}.{4}\spa4.5\ldots\spa{n}.1 \,t_{3,j+3}\,t_{2,j+3}} \,,
\cr
\hat h_{n,r}\; = \; & (-1)^n \,
\hat g_{n,n-r+2}\bigl\vert_{(123..n)\to(321n..4)}\,.
\cr}
\equn\label{ds1}
$$
This expression for the $\NeqOne$ chiral amplitudes agrees with
the one first given in ref.~\cite{BBDPSQCD}.

The expression for the cut-constructible parts of the scalar amplitude
have a very similar form apart from the appearance of $\Lzz$ functions,
$$
\eqalign{
&A_n^{[0]}(1^-,2^-,3^-,4^+,5^+,\cdots, n^+)\; = \; \frac{1}{3}\,A_{n}^{\,\NeqOne\ {\rm chiral}}(1^-,2^-,3^-,4^+,5^+,\cdots, n^+) \cr
&\hspace{2cm}-{i \over 3}
\sum_{r=4}^{n-1}\, \hat d_{n,r}\,
{    \Lzz [ t_{3,r} / t_{2,r} ] \over t_{2,r}^3  }
-{i \over 3}
\sum_{r=4}^{n-2}\, \hat g_{n,r}\,
{    \Lzz [ t_{2,r} / t_{2,r+1} ] \over t_{2,r+1}^3  }
-{i \over 3}
\sum_{r=4}^{n-2}\ \hat h_{n,r}
{    \Lzz [ t_{3,r} / t_{3,r+1} ] \over t_{3,r+1}^3  }
\cr
&  \hskip 2 cm + \hbox{rational}\,, \cr
}
\equn
$$
\noindent
where in this case,
$${
\eqalign{
\hat d_{n,r}\ =\ &
{  \la 3 | {K}_{3,r}{k}_{2}  | 1 \ra \,
  \la 3 | {k}_{2}{K}_{2,r}  | 1 \ra\,
\la 3 | {K}_{3,r}\big[k_2,{K}_{2,r}\big]{K}_{2,r}  | 1 \ra
\over
 [ 2 | {K}_{2,r}  | r \ra
 [ 2 | {K}_{2,r}  | {r+1} \ra\, \spa3.4\ldots \spa{r-1}.{r}
\spa{r+1}.{r+2}\ldots   \spa{n}.1} \,,
\cr
\hat g_{n,r}\; = \; &
\sum_{j=1}^{r-3}
{\la 3|K_{3,j+3} K_{2,j+3}P|1\ra\,
  \la 3|K_{3,j+3} K_{2,j+3}\tilde P |1\ra\,
\la 3|K_{3,j+3}K_{2,j+3}\big[k_{r+1},K_{2,r}\big]|1\ra
\spa{j+3}.{j+4}
\over
[2|K_{2,j+3}|j+3\ra
[2|K_{2,j+3}|j+4\ra\,
\spa{3}.{4}\spa4.5\ldots\spa{n}.1 \,t_{3,j+3}\,t_{2,j+3}} \,,
\cr
\hat h_{n,r}\; = \; & (-1)^n\, 
\hat g_{n,n-r+2}\bigl\vert_{(123..n)\to(321n..4)}\,,
\cr}}
\equn\label{ds0}
$$
and $P=k_{r+1} K_{r+1,1}$ and $\tilde P=K_{r+1,1}k_{r+1}$\,.

Note that for the $\NeqOne$ case the infrared-singularities lie
entirely within the $\Kz$ functions with the expected result,
$$
A^{\; \NeqOne}(1^-,2^-,3^-,4^+,5^+,\cdots, n^+)|_{\eps^{-1}}
\; = \;
{1 \over \eps}
\times
\Atree(1^-,2^-,3^-,4^+,5^+,\cdots, n^+)\,.
\equn
$$


\subsection{Obtaining the NMHV Coefficients from General Formula}

In this subsection we will show how to obtain the results for the NMHV
amplitudes using our generic formula for $T_{P_L,P_R}$ given in
\eqn{generalT}. A coefficient of an ${\rm L}_i$ function will be of
the form
$$
c_{n,3}^{m,r}\;=\;\sum_{P_L,P_R} T_{P_L,P_R}\,.
\equn\label{sumtdg}
$$
If the amplitude has three adjacent negative helicities they can
only be positioned in two intrinsically different ways. In the
first case, the massless leg is $2^-$ and the massive legs are
$r+1^+,\cdots n^+,1^-$ and $3^-,4^+,\cdots r^+$, {\it i.e.} the
coefficients $c_{n,3}^{2,r}$\,. In the second case,
the massless leg has positive helicity and one massive leg contains
two negative helicities and the other one, {\it i.e.} the coefficients
$\bar c_{n,1}^{\bar m,3}$ and $\bar c_{n,2}^{\bar m,3}$\,.

Consider the first case. In this case the paths are degenerate
being from $\alpha_1=3$\,, $\rho_1=r$ to $\alpha_1=3,\,\rho_2=4$ for the right
corner and  from $\beta_1=1,\,\sigma_1=r+1$ to $\beta_1=1$ to $\sigma_2=n$
on the left.
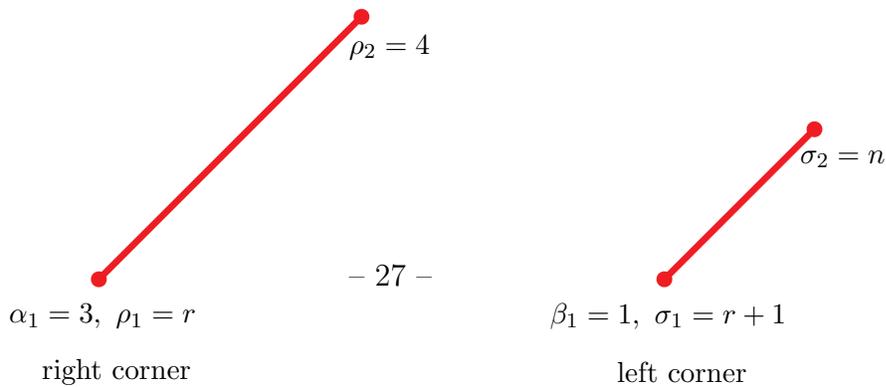
\begin{center}
\vspace{-4.3cm}
\begin{picture}(100,160)(120,0)
\rotatebox{-45}{
\Line(120,40)(120,180)
\SetColor{Red}
\SetWidth{2.5}
\Line(120,40)(120,180)
\Text(130,30)[]{\small \rotatebox{45}{
$\alpha_1=3,\,\,\rho_1=r$}}
\Text(150,20)[]{\small \rotatebox{45}{right corner}}
\Vertex(120,40){3}
\Vertex(120,180){3}
\Text(136,180)[]{\small \rotatebox{45}{$\rho_{2}=4$}}
}\end{picture}
\begin{picture}(100,160)(10,00)
\rotatebox{-45}{
\Line(120,40)(120,120)
\SetColor{Red}
\SetWidth{2.5}
\Line(120,40)(120,120)
\Text(130,30)[]{\small \rotatebox{45}{
$\beta_1=1,\,\,\sigma_1=r+1$}}
\Text(150,20)[]{\small \rotatebox{45}{left corner}}
\Vertex(120,40){3}
\Vertex(120,120){3}
\Text(136,120)[]{\small \rotatebox{45}{$\sigma_{2}=n$}}
}\end{picture}
\vspace{3.2cm}
\end{center}

\noindent
The sum in eq.~(\ref{sumtdg}) then just becomes a single term, which
can be computed along the lines of the previous section.  The explicit
expressions for $c_{n,3}^{2,r}$ are given above in eq.~(\ref{ds1}) or
in eq.~(\ref{ds0}) for the two choices for $P$ and $\tilde P$\,, depending
on whether a scalar or $\NeqOne$ multiplet circulate in the loop.

Let us consider now the second case. Both coefficients $\bar
c_{n,3}^{\bar m,1}$ as well as $\bar c_{n,3}^{\bar m,2}$ are related
to $c_{n,n-3}^{m-3,n-2}$ by parity and relabelling. We consider the
coefficient $\bar c_{n,3}^{\bar m,1}$ in more detail.  The left
massive cluster of legs of the corresponding triangle
is $(2^-,3^-,4^+,\dots,(m-1)^+)$ with $m-4$ positive
helicities. This coefficient is related by conjugation and relabelling
to
$$
\bar c_{n,3}^{\bar m,1}\;=\;
c_{n,n-3}^{m-3,n-2}\big|_{(123..n)\rightarrow (456...3),{\rm parity-flip}}\,.
\equn
$$
The coefficient $c_{n,n-3}^{m-3,n-2}$ can be computed along the lines
of the previous section and gives two kinds of contributions.
For both
the right corner contributes with $\alpha_0=m-2$\,, $\alpha_1=n-3$ and
$\rho_1=n-2$\,, which corresponds to a single path as
displayed below.
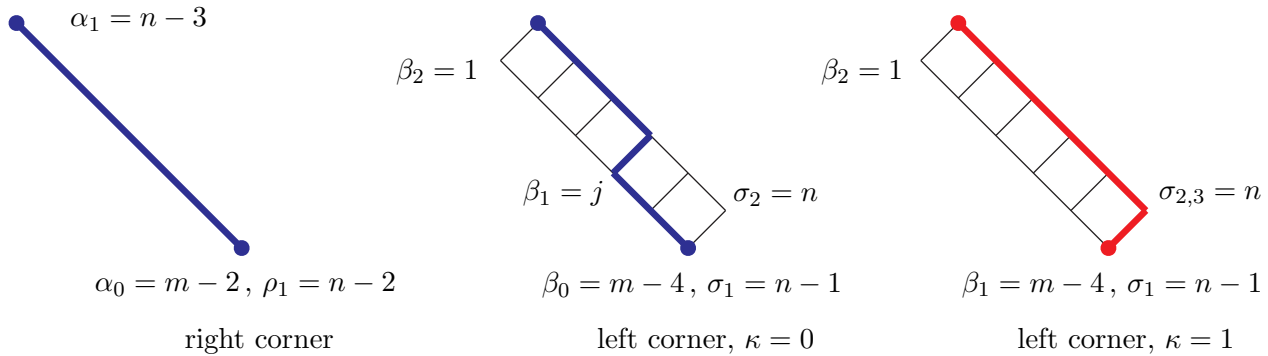
\begin{center}
\vspace{-4.2cm}
\hspace{-0.8cm}
\begin{picture}(100,160)(95,0)
\rotatebox{-45}{
\Line(0,40)(120,40)
\SetColor{Blue}
\SetWidth{2.5}
\Line(0,40)(120,40)
\Text(130,30)[]{\small \rotatebox{45}{
$\alpha_0=m-2\,,\,\rho_1=n-2$}}
\Vertex(120,40){3}
\Vertex(0,40){3}
\Text(42,84)[]{\small \rotatebox{45}{
$\alpha_1=n-3$\quad\quad\quad}}
\Text(150,20)[]{\small \rotatebox{45}{right corner}}}
\end{picture}
\begin{picture}(100,160)(30,0)
\rotatebox{-45}{
\Line(20,40)(120,40)
\Line(20,60)(120,60)
\Line(120,40)(120,60)
\Line(100,40)(100,60)
\Line(80,40)(80,60)
\Line(60,40)(60,60)
\Line(40,40)(40,60)
\Line(20,40)(20,60)
\SetColor{Blue}
\SetWidth{2.5}
\Line(120,40)(80,40)
\Line(80,40)(80,60)
\Line(80,60)(20,60)
\Text(130,30)[]{\small \rotatebox{45}{
$\beta_0=m-4\,,\,\sigma_1=n-1$}}
\Vertex(120,40){3}
\Vertex(20,60){3}
\Text(75,24)[]{\small \rotatebox{45}{
$\beta_1=j\quad$}}
\Text(10,24)[]{\small \rotatebox{45}{
$\beta_2=1\quad\,\,$}}
\Text(125,72)[]{\small \rotatebox{45}{
$\quad\sigma_2=n$}}
\Text(150,20)[]{\small \rotatebox{45}{left corner, $\kappa=0$}}}
\end{picture}
\begin{picture}(100,160)(-25,0)
\rotatebox{-45}{
\Line(20,40)(120,40)
\Line(20,60)(120,60)
\Line(120,40)(120,60)
\Line(100,40)(100,60)
\Line(80,40)(80,60)
\Line(60,40)(60,60)
\Line(40,40)(40,60)
\Line(20,40)(20,60)
\SetColor{Red}
\SetWidth{2.5}
\Line(120,40)(120,60)
\Line(120,60)(20,60)
\Text(130,30)[]{\small \rotatebox{45}{
$\beta_1=m-4\,,\,\sigma_1=n-1$}}
\Vertex(120,40){3}
\Vertex(20,60){3}
\Text(10,24)[]{\small \rotatebox{45}{
$\beta_2=1\quad$}}
\Text(125,72)[]{\small \rotatebox{45}{
$\quad\quad\sigma_{2,3}=n$}}
\Text(150,20)[]{\small \rotatebox{45}{left corner, $\kappa=1$}}}
\end{picture}
\vspace{3.2cm}
\end{center}
For the left corner we have two cases which have to be distinguished.
The first one comes from $\beta_0=m-4$\,, $\beta_1=j$\,, $\beta_2=1$\,,
$\sigma_1=n-1$ and $\sigma_2=n$\,. One finds a term for the values of
$j=1,...,m-5$\,. The other contribution comes from the single argument
$\beta_1=m-4$\,, $\beta_2=1$\,, $\sigma_1=n-1$ and $\sigma_2=n$\,.
The sum over all paths that contribute to $\bar c_{n,3}^{\bar m,1}$ is thus
$$
\eqalign{
\bar c_{n,3}^{\bar m,1}&\;=\;-i\hat g_{n,m-1}\,,\cr
\bar c_{n,3}^{\bar m,2}&\;=\;-i\hat h_{n,m-1}\,,
}
\equn
$$ where $\hat g_{n,r}$ and $\hat g_{n,r}$ are given explicitly in
eq.~(\ref{ds1}) or eq.~(\ref{ds0}) for the two values of $P$ and
$\tilde P$ corresponding to an $\NeqOne$ multiplet or scalar in the
loop.  Notice that the sum over $j$ in the expression for $\hat
g_{n,r}$ represents the contributions for the two cases $\kappa=0$
and $\kappa=1$ together.  It is striking that the computation of
these coefficients is no more difficult than that of a tree-level
calculation.


\section{Conclusion}

In the endeavour to compute the perturbative interactions within gauge
theories, techniques which make direct use of previously computed
amplitudes to generate new ones are usually to be preferred.  The
loop-level unitarity method~\cite{BDDKa,BDDKb} or tree-level on-shell
recursion relations~\cite{TreeRecursion,BCFW} are examples of this.
In this paper we developed an on-shell recursive method for obtaining
coefficients of integral functions from previously calculated ones.
We presented sufficiency criteria for avoiding spurious singularities
and boundary terms in the recursion.  These criteria apply to a large
class of coefficients.

We illustrated this with a non-trivial example, obtaining the complete
$n$-gluon scattering with an $\NeqOne$ chiral multiplet circulating in
the loop and where the external gluon helicities are ``split'',
$A_n(1^-, 2^-, \ldots, j^-, (j+1)^+, \ldots, n^+)$\,.  For this same
helicity configuration with a scalar circulating in the loop we also
obtained all logarithmic terms in the amplitudes.  Our starting point
in the recursion were the cut containing parts of the five-point MHV
amplitudes obtained in ref.~\cite{FiveGluon}.  With the supersymmetric
decomposition, the contributions we have computed are pieces of QCD amplitudes.

A particularly simple subset of the amplitudes obtained here is
the NMHV case with three colour-adjacent negative helicities and the
rest positive.
In this case, the $\NeqOne$ amplitudes agree with previous
computations~\cite{BBBDDNEq1,BrittoSQCD}, providing a non-trivial check
on our methods.  The cut parts of the scalar loop contributions are
new.  Since the NMHV $n$-point amplitudes with an $\NeqFour$ multiplet
circulating in the loop are also known~\cite{BDKn}, the only missing
pieces for complete QCD amplitudes are the rational function terms.

It would be very interesting to apply the ideas discussed in this
paper to obtain other coefficients of integral functions appearing in
one-loop QCD amplitudes.  A new means of dealing with the rational
function terms of loop amplitudes has also been given
recently~\cite{BDKSix}.  We may look forward to many new multi-parton
one-loop amplitudes for use in collider physics.


\acknowledgments

We thank Andi Brandhuber, Nigel Glover, Warren Perkins and
Gabriele Travaglini and especially Lance Dixon and David Kosower
for many helpful discussions and observations. This research was supported
in part by the US Department of Energy under grant
DE--FG03--91ER40662 and in part by the EPSRC and the
PPARC of the UK.



\end{document}